\documentclass[preprint,12pt]{elsarticle}
\usepackage[section]{placeins}

\usepackage{makeidx} 
\usepackage{epsfig,wrapfig}
\usepackage{amsmath,amscd,stmaryrd}
\usepackage{amssymb,amsmath,latexsym}
\usepackage{subfigure,wrapfig}
\usepackage{colortbl}
\usepackage{multirow}
\usepackage{multicol}
\usepackage{longtable}
\usepackage[table]{xcolor}
\usepackage{url}
\usepackage[T1]{fontenc}
\usepackage{array,multirow}
\usepackage{caption}
\usepackage[utf8]{inputenc}

\usepackage{appendix}

\usepackage{hyperref}

\usepackage{url}
\usepackage{listings} 
\usepackage[scaled=.75]{beramono}

\usepackage{utilities}

\usepackage{enumitem}

\usepackage{enumitem}

\definecolor{javared}{rgb}{0.6,0,0} 
\definecolor{javagreen}{rgb}{0.25,0.5,0.35} 
\definecolor{javapurple}{rgb}{0.5,0,0.35} 
\definecolor{javadocblue}{rgb}{0.25,0.35,0.75} 

\lstdefinestyle{JavaColors}
{
	captionpos=b,
	basicstyle=\ttfamily\footnotesize,
	keywordstyle=\color{javapurple}\bfseries,
       xleftmargin=15pt,
	stringstyle=\color{javared},
	commentstyle=\color{javagreen},
       numberstyle=\tiny,
	morecomment=[s][\color{javadocblue}]{/**}{*/},
      numbers=left,
	nolol=true,
	tabsize=4,
	showspaces=false,
	showstringspaces=false
      numberbychapter=false,
	numbersep=5pt,
      columns=[l]fullflexible,
}

\lstdefinestyle{intable}{
        basicstyle=\scriptsize\sffamily,
        tabsize=1,
        xleftmargin=7pt,
        numbersep=5pt,
        breaklines=true,
        morecomment=[l]{\#},
        keywordstyle=[2]\ttfamily,
        postbreak=\mbox{{\color{red}\tiny$\hookrightarrow$}},
        aboveskip=-0.6\baselineskip,
        belowskip=-0.8\baselineskip,
}

\lstdefinestyle{XtextColors}
{
	basicstyle=\ttfamily\small,
	keywordstyle=\color{javapurple},
	stringstyle=\color{javared},
	commentstyle=\color{javagreen},
	morecomment=[s][\color{javadocblue}]{/**}{*/},
	tabsize=4,
	showspaces=false,
	showstringspaces=false
}

\lstdefinelanguage{Xtext}
{
	morekeywords={returns, name},
	sensitive=true,
	morecomment=[l]{//},
	morecomment=[s]{/*}{*/},
	morestring=[b]",
}

\lstdefinelanguage{OCLUSE}
{
keywordstyle=\color{black}\bfseries,
keywords = [2]{ANN, MOD, TYPE, ANN1, MOD1, MODS1, TYPE1, ANNn, MODn, MODSn, TYPEn, ANNi, MODi, MODSi,TYPEi},
morekeywords={association,between,role,end,context, inv, implies,or,and,xor,if,then,else,endif,not,Sequence,exists, not, forAll},
sensitive=true,
morecomment=[l]{--},
morestring=[b]',
showstringspaces=false,
xleftmargin=3pt,
aboveskip=0pt,
belowskip=3pt,
}

\newcommand{\code}[1] {{\footnotesize\sffamily #1}}

\lstdefinelanguage{Annotation}
{
	morekeywords={public, private, protected, package, 
				int, long, short, char, byte, boolean, float, double, String, Class,
				extends, require, forbid, or, at, and, annotation,
				enum, interface, method, constructor, field, class, all,
				static, final, abstract,
				runtime, classfile, source},
      keywords = [2]{ANN, MOD, TYPE, ANN1, MODS1, TYPE1, ANNi, MODSi, TYPEi, ANNn, MODSn, TYPEn},
	sensitive=true,
	morecomment=[l]{//},
	morecomment=[s]{/*}{*/},
	morestring=[b]",
}

\lstdefinelanguage{EBNF}
{
	morekeywords={public, private, protected, package, 
		int, long, short, char, byte, boolean, float, double, String, Class,
		extends, require, forbid, or, at, and, annotation,
		enum, interface, method, constructor, field, class, all,
		static, final, abstract,
		runtime, classfile, source, [},
	sensitive=true,
	morecomment=[l]{//},
	morecomment=[s]{/*}{*/},
	morestring=[b]",
}

\lstdefinestyle{EBNF}
{
	captionpos=b,
	basicstyle=\ttfamily\footnotesize,
	keywordstyle=\bfseries,
  xleftmargin=15pt,
	stringstyle=\bfseries,
	commentstyle=\color{javagreen},
       numberstyle=\tiny,
	morecomment=[s][\color{javadocblue}]{/**}{*/},
      numbers=left,
	nolol=true,
	tabsize=4,
	showspaces=false,
	showstringspaces=false
      numberbychapter=false,
	numbersep=5pt,
      columns=[l]fullflexible,
}

\journal{ Computer Languages, Systems and Structures}

\begin{document}
\begin{frontmatter}

\title{Ann: A domain-specific language for the effective design and validation of Java annotations}

\author{Irene Córdoba-Sánchez}
\ead{icordoba@fi.upm.es}
\author{Juan de Lara}
\ead{Juan.deLara@uam.es}

\address{Computer Science Department, Universidad Aut\'onoma de Madrid (Spain)}
\begin{abstract}
This paper describes a new modelling language for the effective design and validation of Java annotations.
Since their inclusion in the 5th edition of Java, annotations have grown from a useful tool 
for the addition of meta-data to play a central role in many popular software projects. 
Usually they are not conceived in isolation, but in groups, with dependency and integrity constraints between them.
However, the native support provided by Java for expressing this design is very limited. 

To overcome its deficiencies and make explicit the rich conceptual model which lies behind a 
set of annotations, we propose a domain-specific modelling language. 
The proposal has been implemented as an Eclipse plug-in, including an editor and an
integrated code generator that synthesises annotation processors. 
The environment also integrates a model finder, able to detect unsatisfiable constraints 
between different annotations, and to provide examples of correct annotation usages for validation.
The language
has been tested using a real set of annotations from the Java Persistence
API (JPA). { Within this subset we have found enough rich semantics expressible with Ann and omitted nowadays by the Java language, which shows the benefits of Ann in a relevant field of application.}
\end{abstract}

\begin{keyword}
Model Driven Engineering, Domain-Specific Languages, Code generation, Java annotations, Model Finders
\end{keyword}

\end{frontmatter}

\section{Introduction}\label{intro}
In 2004 the possibility of adding custom meta-data to  programs was added to the Java language in the form of \emph{annotations}\footnote{\url{http://docs.oracle.com/javase/1.5.0/docs/guide/language/annotations.html}}. Predefined annotations were available previously for very specific tasks, however,  the huge amount of boilerplate code that many Application Programming Interfaces (APIs) required motivated their establishment as another general tool in the language. Schildt~\cite{Schildt:javaann} suggests other reasons as motivation to the appearance of annotations in Java: the increasingly growing tendency of including the meta-data associated with a program
within the program itself instead of keeping it in separate files; and the pressure
from other programming languages which already included similar features, like C\#.

Since their introduction in the language, annotations have become a success and
are widely used in many important projects within the software development scene. We find them
in frameworks like Seam\footnote{\url{http://docs.jboss.org/seam/latest/reference/html/annotations.html}} and Spring\footnote{\url{http://spring.io/}}, in the Object Relation Mapping of Hibernate\footnote{\url{http://hibernate.org/orm/}}, and also in proper Sun Java standards such as the Java Persistence API (JPA)\footnote{\url{http://www.oracle.com/technetwork/java/javaee/tech/persistence-jsp-140049.html}}.

However, despite this success, the native support that Java provides for their construction is very poor. 
{This is so as Java annotations are not defined using a specialized syntax, but reusing the syntax to create interfaces.
This lack of specialized syntax greatly limits the ability to specify the elements where an annotation can be 
placed and further correctness conditions.}
Moreover, annotations are rarely conceived in an isolated way; instead they are usually part of
a set with dependencies and integrity constraints. Currently there is no effective way in Java
for making explicit the constraints underlying a set of annotations at design time, and validate that they
are not conflicting. Instead, the usual
path taken to overcome these deficiencies is to develop an extension to the Java compiler
(called annotation processor) to ensure that such constraints are complied with, and rely on extensive manual 
testing of such processor.

As a first step towards the alleviation of this situation, we propose {\em Ann}, a textual Domain-Specific Language 
(DSL) \cite{Voelter} aiming to provide a
more expressive and suitable syntactic support for the design of sets of annotations and their associated integrity constraints. We have developed an Integrated Development Environment (IDE) as an Eclipse plug-in, which integrates seamlessly with the
Java IDE. The environment includes a code generator to translate the design and constraints expressed using {\em Ann} into Java code, which can be fully integrated into projects in such language. Moreover, {\em Ann} makes use of a constraint solver over models (a model finder~\cite{USE}), which is able to detect whether the constraints posed by a set of annotations are 
unsatisfiable, and provide examples (annotated class mock-ups) of usages of the annotations. 
These examples could be used by designers to validate whether the encoded integrity
constraints defined with {\em Ann} are according to their intentions.
\emph{Ann} has been tested using a real set of annotations from JPA,
demonstrating that it can capture a wide set of the constraints in its specification, and showing advantages with respect to other approaches with similar goals in the state of the art. More information and the source code
of the project can be found at \url{http://github.com/irenecordoba/Ann}.

This paper is an extended version of~\cite{Ann,AnnTFG}, where we have added validation mechanisms based on constraint solving, and integrated such mechanisms with the environment. Additionally, all sections have been enlarged with additional explanations and more details.

The rest of the paper is organised as follows. 
Section~\ref{related} analyses related work. Section \ref{javaann} gives a more detailed overview on the current limitations of
Java annotations. Section~\ref{mdd} introduces the main concepts of Model-Driven Engineering (MDE) and the different choices when building a DSL. 
Section~\ref{overview} provides an overview of our approach.
Section \ref{dslann} describes the proposed DSL, {\em Ann}. Section~\ref{validation} explains our approach to validate annotations. 
Section \ref{usecase} contains the evaluation of {\em Ann}, including a detailed real case study and an evaluation of the efficiency of the model finder for
annotation validation. Finally, section \ref{conclusions} summarises the conclusions and 
future development. Two appendices detail the description of the textual concrete syntax of Ann, and the generated OCL invariants for the JPA case study.

\section{Related research}\label{related}
Some research has been made in order to improve and expand the functionality of Java annotations.
For example,  Phillips in~\cite{Phillips:2009:CMA:1639950.1640005} aims at conciliating object oriented principles
with the design of annotations by the introduction of a new one: {\tt composite}. With it, he manages to
support composition, allowing encapsulation and polymorphism of annotations.

A Java extension, {\tt @Java}, is proposed by Cazzola and Vacchi~\cite{Cazzola20142}
in order to expand the range of application of an annotation to code blocks and expressions, although some
improvement in this respect has also been made natively in the latest version of Java\footnote{Java 8 at the time of writing}.

The expressiveness limitations of Java annotations are recognised in~\cite{ClarkSW08}, where a proposal
is made to embed DSLs into Java, with a more natural and flexible syntax. JUMP~\cite{JUMP} is a tool
to reverse engineer Java programs (with annotations) into profiled UML class diagrams.

Although the aforementioned approaches expand the features of Java annotations, they do not facilitate their design,
nor address the 
limitations with respect to expressing integrity constraints within an annotation or between the annotations in
an annotation set, which is the main goal of our work. 

Just a few works are aimed at improving the design of annotations.
Darwin~\cite{Darwin:annabot} suggests a DSL, called AnnaBot, based on \emph{claims} about 
a set of existing annotations, with a concrete syntax very similar to Java. With these \emph{claims}, interdependency
constraints can be expressed within a set of annotations. However, there is no possibility of characterising the
valid targets of an annotation type (i.e., the valid elements in a program where the annotation can be placed). 
Moreover, no improvement is made with respect to the syntax for defining 
annotations in Java, given 
its heavy focus on existing sets of annotations and constraints between them, and not on isolated ones. Finally,
the approach uses reflection to check the statements of its \emph{claims}, which could and should be avoided.

Another approach is AVal~\cite{noguera:aval}, a set of meta-annotations (annotations which are placed at other annotations) to add integrity constraints at the definition of the annotation type. 
The drawback of this approach is that its expressiveness is rather restricted, given the limited 
flexibility which the structural characteristics of meta-annotations provide.

Pluggable type systems~\cite{Bracha04}, provide a way to support improved analysis of programs by ensuring stronger type checking.
Implementation of these systems exist for Java, like the JavaCOP framework~\cite{JavaCOP}. Java 8 improves the support for such systems
via type annotations. Pluggable type systems provide a customized semantics to sets of programming elements via the use of annotation sets. 
However, they normally do not provide a unified way to describe and validate the syntax and integrity constraints of a set of annotations.

Regarding constraint validation, constraint solving strategies are extensively used in connection with program testing~\cite{DeMilloO91}. Many times,
constraint solving is used as a means to generate interesting test data for programs, perhaps derived from specifications~\cite{GuerraS15}, or making the
program execute a certain path~\cite{ISSRE}. In~\cite{IslamC14} constraint solving is used to generate both test cases and
mock-up classes, considering advanced features like reflection and annotations. Hence, while they can generate
mock-up programs with annotations, they need to derive the needed constraints
from the analysis of reflective Java code (i.e., at runtime), with the consequent loss of precision and drawbacks of runtime checking. 
Instead, for our purposes, it would be more desirable to have a language making explicit the annotation constraints at design time, 
which then can be analysed before such annotations are used.

Hence, as we have observed, there is currently a need for: (i) better syntactical support for the specification of Java annotations;
(ii) explicit, high-level means to describe constraints for an annotation (e.g., regarding its valid targets), and between a set
of annotations (e.g., expressing their dependencies); and (iii) ways to analyse such constraints to find inconsistencies, at design time.
For this purpose, in the rest of the paper, we describe Ann, a DSL directed to describe both the
syntax and well-formedness constraints of annotation families, to validate the correctness of the annotation constraints and
to make explicit the design of such annotation set, allowing their immediate use on Java projects thanks 
to the code generation facility.

\section{Java annotations}\label{javaann}
To help understanding the current limitations of Java annotations, in this section we describe how 
they are defined in Java (subsection~\ref{sec:jann_def}), their usage and limitations
(subsection \ref{sec:ann_usage}) and how their correct use is checked (subsection~\ref{sec:ann_proc}).

\subsection{Defining Java annotations}\label{sec:jann_def}

Java annotations do not constitute a type of their own. Instead, they are defined as \emph{special}
interfaces. There are many differences between annotations and interfaces, however we will only review
those necessary to understand the design of Ann.

Listing~\ref{java_def} shows an example of the definition and usage of a simple annotation called \code{Person}.
Such annotation definition is called an \emph{annnotation type}.
As it can be noticed, the special nature of annotations is indicated by the \texttt{@} character 
before the \texttt{interface} keyword (line 7). The zero-argument methods inside the container (lines 7-9)
are the \emph{fields} (the parameters) of the annotation. To assign a default value to those \emph{fields},
the keyword \texttt{default} must be used.
An annotation can have an arbitrary number of fields, which can be of primitive type, \code{Class}, \code{String}, an annotation
type, or an array of the previous types.

\begin{figure}[!t]
\begin{lstlisting}[
	style=JavaColors, 
	language=Java, 
%	caption=Annotation \texttt{Person} defined in Java.,
%	label=java_def
]
package examples;

import java.lang.annotation.*;

@Target(ElementType.TYPE)
public @interface Person {
    String name() default "Mary";
    int age() default 21;
    float weight() default 52.3f;
}
\end{lstlisting}
\captionof{lstlisting}{Annotation \texttt{Person} defined in Java.}
\label{java_def}
\end{figure}

Since the goal of annotations is to add meta-data to Java programs by being placed at certain elements, it is important to
know which elements are eligible as their \emph{targets}. Annotations can be employed in the declaration of many constructions
in Java; however with Ann we have focused on the most usual ones, namely types (classes, interfaces,
enumerates, annotations), constructors, methods and fields\footnote{{Ann was started before the official release of Java 8,
and hence some of its feature regarding annotations, like the possibility of defining type uses as targets is not currently supported, 
but left for future work.}}.

Another important characteristic of annotations is that they can have three different levels of \emph{retention},
which depends on the phase where they will be used: in the source code (they are discarded by the compiler); compiled but 
ignored by the Java Virtual Machine (JVM); and compiled and read by the JVM when the type that contains them is loaded.
The last ones are accessible by the Java Reflection API at runtime, which can check the values in their fields.

Both the targets allowed for an annotation and its desired level of retention can be specified at design time by using
the standard Java meta-annotations \texttt{Target} and \texttt{Retention}, respectively. As we will see in the next
subsection, the former is very poor regarding its expressive power with respect to real and common use cases where
annotations are used.

\subsection{Annotation usage and limitations}\label{sec:ann_usage}

Listing~\ref{java_usage} shows the use of the defined {\tt Person} annotation.
Annotations are considered as modifiers when using them on a target in Java. This is why, although
in Listing \ref{java_usage}, line 3, annotation \texttt{Person} appears above the class, it could perfectly
be merged with the rest of the modifiers, but the former is the usual syntax.

Pairs \emph{\texttt{key = value}} are used in order to specify the values of the fields of the annotation.
It is mandatory to set a value for all the fields that do not have a default value predefined on the annotation type,
and the order of the fields does not matter.

\begin{figure}[!t]
\begin{lstlisting}[
style=JavaColors,
	language=Java, 
%	caption=Annotation \texttt{Person} defined in Java.
]
import examples.Person;

@Person(name="Peter", age=43)
class Filter {
...
}
\end{lstlisting}
\captionof{lstlisting}{Usage of the annotation \texttt{Person}.}
\label{java_usage}
\end{figure}

Traditionally only one instance of a particular annotation type could annotate a target; however with 
Java 8 it is possible to use several such instances if the corresponding annotation
type is properly marked on its definition.

Line 5 of Listing \ref{java_def} shows another example of an annotation being used: \code{Target}. In this case,
the value is directly specified because the annotation has only one field and it is named \texttt{value}. By using the value {\tt TYPE} of the
enumeration {\tt ElementType}, {\tt Person} is restricted to be applied to classes, interfaces (including annotation types) and 
enumerations. However, there is no way to e.g., restrict its applicability to classes only.

We have presented a very simple example of an annotation type, but if we take a look at the JPA documentation,
particularly the extensively used \texttt{Entity} annotation, we find that it can 
only be applied to classes meeting the following more elaborated requirements\footnote{\url{http://docs.oracle.com/javaee/7/api/}}:
\begin{itemize}
	\item They must have a {\em public} or {\em protected constructor}.
	\item They must not be {\em final}.
	\item They must not have {\em any final method}.
	\item Their {\em persistent fields} must be declared {\em private}, {\em protected} or {\em package-private}.
\end{itemize}

None of these constraints can be expressed nowadays with the syntax available for the definition of annotation types.

What is more, when designing annotation sets, it is common to have constraints involving several annotations, because
the annotations are usually inter-related. For
example, the JPA annotation {\tt Id} is only allowed in attributes within classes annotated with {\tt Entity}.
We call such constraints the static semantics or integrity constraints of an annotation, or an annotation set. 
Given a large and complex set of requirements for an annotation set, it is easy to make mistakes, by requiring
conflicting features from the different annotations, specially at design time. Just for the sake of illustration, if we require {\tt Id} to 
be applicable on a public attribute, then it would make the {\tt Id} annotation to be in conflict with {\tt Entity}
(as the latter requires non-public attributes), and hence inapplicable.

Therefore, what can be done to ensure the compliance of such outlined constraints and ensure their validity? 
The only remaining choices are to write a 
guiding comment for its use and signal an error at runtime. In addition, it is possible to develop extensions to the 
Java compiler, known as annotation processors, and rely on their extensive manual testing to validate that the
annotation requirements are met by their implementation in the processor. In the next subsection we review these annotation processors,
since they are one of the key components of Ann.

\subsection{Annotation processors}\label{sec:ann_proc}

The Java package \texttt{javax.annotation.processing} provides a set of elements for processing annotations at compile
time. An annotation processor is invoked by the compiler, and it can check the annotations attached to any
program element, performing an arbitrary task. Typically, the processor will check the correctness of the annotation
placement (i.e., its static semantics), and may perform further actions (e.g., generating code).
Annotation processing works in rounds. In each round a processor may be required to process a subset of the annotations
found in the source code and the binary files produced in the prior round. If a processor was executed in a given round,
it will be called again in the next rounds. 

Listing \ref{ann_proc} shows the structure of a typical annotation processor. Line 1 specifies the annotation to be checked, 
{\tt Person} in this case. 
The key method of the processor is {\tt process} (lines 5-23), where the elements annotated with the particular annotation
are looked up and checked. If any of them does not satisfy the checks, then an error is raised using the functionality provided by the 
\texttt{processing} package (lines 15-20).

\begin{figure}[t!]
\begin{lstlisting}[
	style=JavaColors, 
	language=Java, 
	caption=Structure of an annotation processor.,
	label=ann_proc
]
@SupportedAnnotationTypes("Person") // annotation to be checked
@SupportedSourceVersion(SourceVersion.RELEASE_6)
public class PersonProcessor extends AbstractProcessor 
{ 
	@Override
	public boolean process(Set<? extends TypeElement> annotations, 
                                   RoundEnvironment objects) 
    {	
        // iterate over all objects to check
        for (Element elt: objects.getElementsAnnotatedWith(Person.class)) 
        {
            // evaluate correct placement of Person annotation for elt
            ...
            // if error
            this.processingEnv.getMessager().printMessage
            (
                Kind.ERROR, 
				"The annotation @Person is disallowed for this location.", 
                elt
            );
        }	
		return true;
	}
}
\end{lstlisting}
\end{figure}

It is important not to confuse annotation processing with reflection. While the former takes place at compile time, the latter
is at execution time. The values of an annotation at a given program element can be checked at execution time via the Java 
Reflection API (if the annotation type is properly marked, as explained in Section \ref{sec:jann_def}), but it has several disadvantages, like an overhead in performance,
the requirement of runtime permission (which may not be granted), and the possibility of breaking object-oriented
abstractions. 

In the context of checking the correctness of annotations, it is more appropriate to do it via annotation processors, because
they can find and signal the errors without the need to execute the program.
However, coding and testing such processors is tedious, cumbersome and error prone. It requires long cycles for coding, installing the processor,
and testing.
Moreover, we believe it would be advantageous to make
explicit the underlying annotation constraints at a higher level, together with the annotation structure. In addition, this would facilitate the analysis 
of annotation conflicts at design time, with no need to install the processor to make those tests.
For this purpose, we have
created {\em Ann}, a DSL to define the structure and integrity constraints of Java annotations, and validate their correctness.

\section{Model-Driven Engineering and Domain Specific Languages}\label{mdd}
For the development of Ann we have followed what is called Model-Driven Engineering (MDE) \cite{Brambilla:mdse,RodriguesdaSilva}, which is characterised
by the use of models as the main component of the development process. A model is a simplified or partial representation of reality, defined
in order to carry out a specific task or reach an agreement on some matter. A great advantage of MDE is that it fills the communication 
gap between the requirements and analysis phase and the implementation phase in a software project.

Modelling languages are extensively used in MDE, and are conceptual tools to describe reality in an explicit way, at some level of abstraction and
from a certain point of view. They are composed and defined by three key elements:
\begin{itemize}
	\item \textbf{Abstract syntax}. It describes the structure of the language and the way in which the different elements can be combined.
	\item \textbf{Concrete syntax}. It describes the particular representation of the modelling language, covering
	features such as the codification or visual appearance, and hence it determines how users visualize or create models. It can be graphical~\cite{Tolvanen,Monaco,JavaCard} or textual~\cite{Voelter,Easytime}.
	\item \textbf{Semantics}. It describes the meaning of the language elements and also the meaning of the different
	ways of combining them.
\end{itemize}

Modelling languages can be classified depending on their domain of application. A Domain-Specific Modelling Language (DSML) is a
modelling language designed for a specific domain or context, with the purpose of easing the task of describing the elements
in such domain. In contrast, General-Purpose Modelling Language (GPML) can be applied to a much broader context. This distinction
is not always easy to draw as it depends on what we consider as a domain (e.g., we could consider the general problem of modelling as a 
specific domain).

\begin{figure}
\centering
	\includegraphics[width = 8cm]{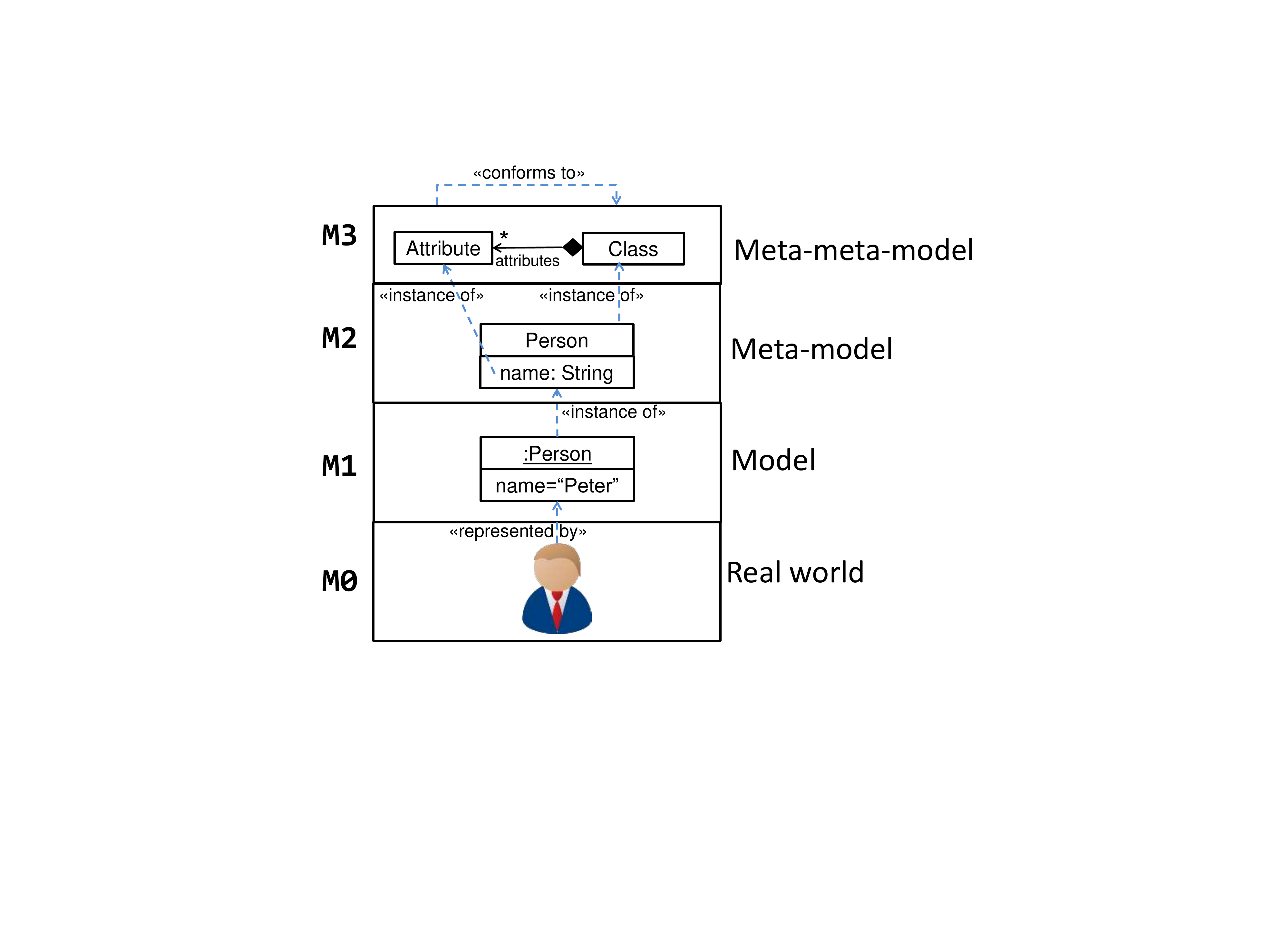}
  \caption{Models, meta-models and meta-meta-models.}
	\label{levels}
\end{figure}

Given that models play a key role in MDE, and they constitute an abstraction of the real world, a natural step is to 
represent them as \emph{instances} of higher levels of abstraction, i.e., higher levels of models or meta-models. Consequently,
using this definition, meta-models describe the set of models considered valid. They define the abstract syntax of a modelling language, since they are a way
of describing all the types of models that can be represented with such language.
We could iterate this abstraction levels and obtain meta-meta-models and so on. However,
in practice, it has been shown that meta-meta-models are enough to describe themselves (see Figure \ref{levels})~\cite{Brambilla:mdse}.

The two main alternatives for defining the semantics of modelling languages are code generation and model interpretation~\cite{HarelR04}. They can
be thought of as the analogous for compilers and interpreters in the case of programming languages, respectively.
A code generator can be thought of as a \emph{model compiler}, that generates executable code from a high level model in order
to create a functional application. This generation of code is usually done by using template languages like Acceleo\footnote{\url{http://www.eclipse.org/acceleo/}}
or the Epsilon Generation Language~\cite{EGL}.
On the other hand, model interpretation is based on implementing a general tool that translates and executes the model 
on the fly.

\begin{figure}
\centering
	\includegraphics[width = 14cm]{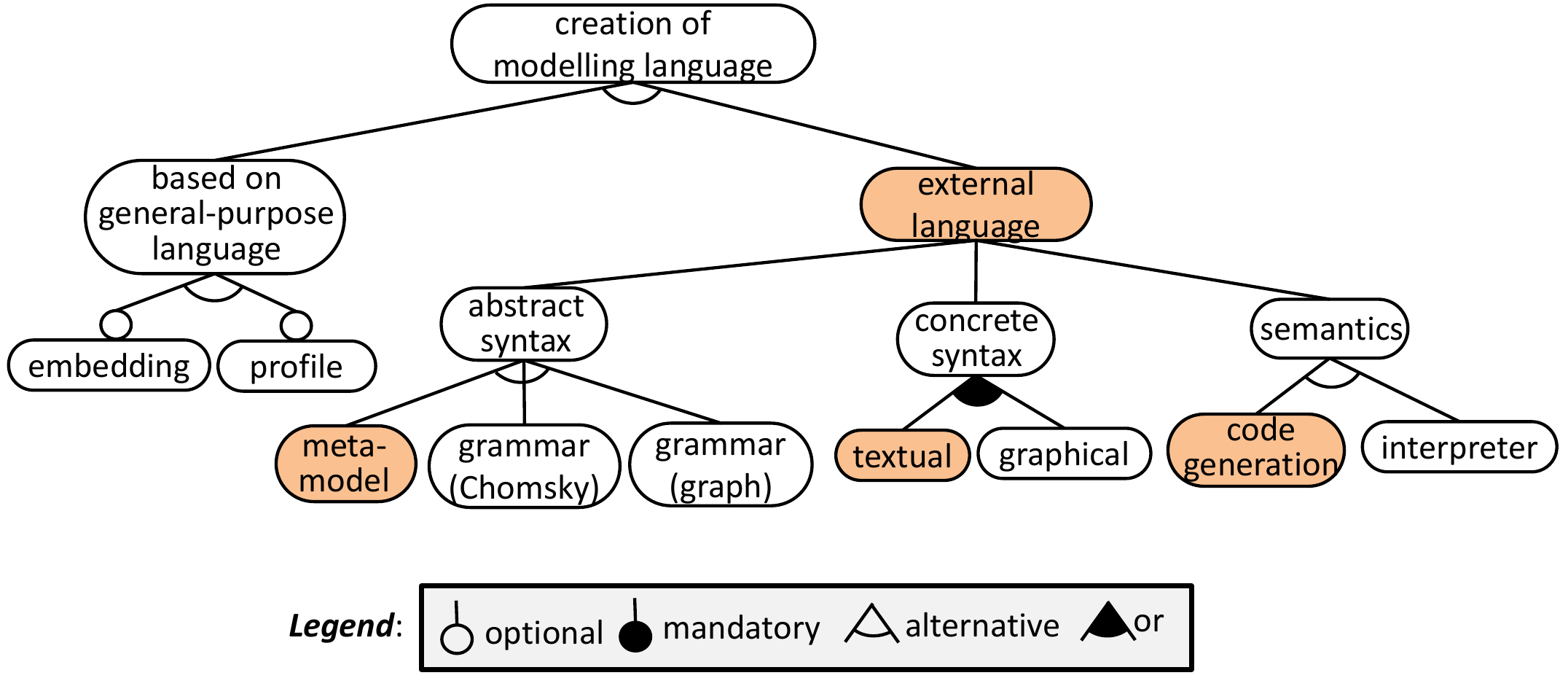}
  \caption{Choices when designing a DSML (Ann choices are shaded).}
	\label{choices}
\end{figure}

Figure~\ref{choices} shows a feature model~\cite{FeatureModel} summarizing the possible choices when designing a modelling language. The diagram is not meant to be exhaustive,
but to gather the most common and typical choices. In the first place, a language can be designed taking
as a basis a GPML, or a general purpose programming language. Using a GPML like the UML, one can design a profile~\cite{profiles}, with stereotypes annotating the
different UML elements and providing domain-specific concepts. This is the approach taken by JUMP~\cite{JUMP}. A DSML can also be embedded in
a general purpose programming language (the so called internal languages). The flexible syntax and dynamic features of languages like Ruby, make them
especially amenable for this task~\cite{CuadradoM09}. Alternatively, one can use a GPML ``as is'', but then domain-specific concepts have to be expressed
as conventions (e.g., naming conventions), programming idioms, or remain at the level of APIs in the case of general purpose programming languages.

{The alternative is the definition of a DSML independent of a base GPML. These are called external DSMLs. To define the abstract syntax, a meta-model
can be used, as previously explained. While this is the standard choice when using MDE, it is not the only alternative. For example, it is possible to define a 
(Chomsky) grammar instead (especially if the language has a textual concrete syntax)~\cite{Voelter}, or a graph-grammar that defines the set of admissible models~\cite{FurstMM15}. Deciding between graphical
and textual concrete syntax is not exclusive, and there are languages featuring both~\cite{GPSS}. Finally, interpretation and code generation are normally alternative choices.
The figure shows in colour the choices made for Ann, whose rationale will be explained in Section~\ref{dslann}.}

\section{Overview of the approach }\label{overview}
Since the goal of Ann is to make explicit the conceptual model behind a set of annotations, 
using a modelling language is a robust choice, because that is what they are precisely designed for. We
have decided to restrict the domain to Java annotations, and that is why we have developed a 
Domain-Specific Language (DSL).

\begin{figure}[h!]
\centering
	\includegraphics[width = 13 cm]{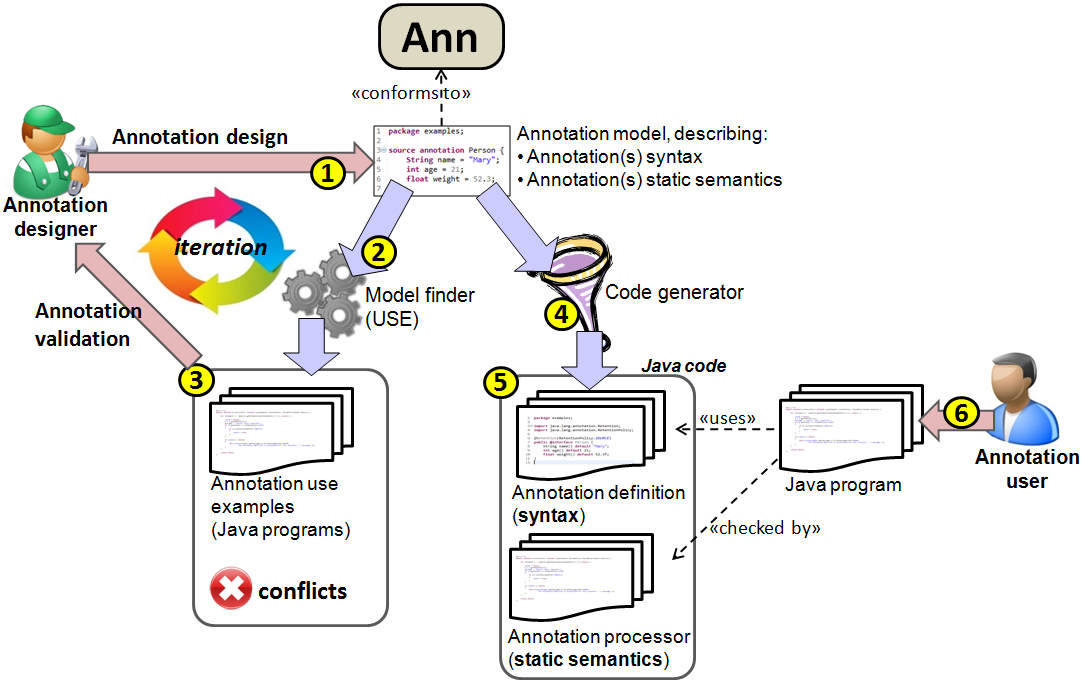}
  \caption{Overview of our approach.}
	\label{fig:overview}
\end{figure}

Figure~\ref{fig:overview} shows the working scheme of our approach to solve the
problems outlined in Section \ref{javaann} by using the DSL. The main idea is to describe 
the syntax and static semantics
of the family of annotations to be built in a declarative way (label 1 in the scheme). 
The Ann DSL provides appropriate
primitives for this task, beyond those natively offered by Java. 

We have incorporated a model finder in our approach, in order to check conflicts between the integrity
constraints of the different annotations. A model finder is a constraint solver over models~\cite{USE}. This way,
its goal is trying to find a model satisfying a number of constraints. In our implementation, we use the USE validator~\cite{USE} (label 2),
a model finder that takes as input a meta-model and a set of invariants expressed in the Object Constraint Language (OCL)\footnote{\url{http://www.omg.org/spec/OCL/}}. As output, it produces a model, a valid instance of the meta-model that satisfies all OCL constraints. In our case, the sought model is a Java program that contains annotations satisfying all the designed integrity constraints. If a model does not exist, then there is some conflict in the annotation set. 
As the sought models are fragments of Java programs (in the form of models) containing the designed annotations (label 3), we can render them as textual Java programs. These could serve designers as a means for validation, as they can check whether the produced example fulfils their requirements and expectations about the designed constraints. Actually, this approach 
is very natural to be used in an iterative process, where the model finder is used to check conflicts and generate examples, and the results of the validation are used to re-design the annotation set, if needed.

Once the annotation design is satisfactory, the designer can make use of a code generator (label 4) that 
produces plain Java files with 
the annotation type definition and the annotation processors for the defined annotation (label 5). Then, the
annotations can be safely used (label 6), because their definition does not contain conflicts, and their correct use in 
Java programs is checked by the generated annotation processors. 

Altogether, using Ann has several advantages, including: (i) it allows to make explicit the structure and integrity constraints
of a set of annotations in a high-level, declarative way; (ii) it provides automatic check of conflicts between constraints at design-time, as well as a generator of annotated example Java programs; and (iii) it automatically produces the annotation
processors to check the correct use of annotations. 

The next section explains the Ann DSL, including its supporting environment,
while the details on how to express Ann constraints in OCL and the interaction with the model finder are left to Section~\ref{validation}.

\section{The Ann DSL}\label{dslann}
Figure \ref{choices} showed
the different options when designing a DSML, showing in colour the design choices for Ann.


First, Ann has been designed as an external language. We could have opted for a UML profile, which could be a sensible choice,
but we preferred a tighter integration with Java programming environment. An internal language within Java was also discarded,
as the Java syntax currently does not offer great flexibility for language embedding.

Given that one of the goals of Ann
is to give a more user-friendly syntax for Java developers defining annotations, mitigating the incoherences that can be found
nowadays in the Java language, a textual concrete syntax has been chosen for it. 
An alternative graphical concrete syntax to facilitate expressing and visualizing the integrity constraints within an annotation set could be interesting,
but is left for future work.
The abstract syntax was designed using a meta-model, which is the standard choice when using MDE.

For the semantics of Ann, code generation has been the adopted solution. While an interpreter was also possible, we opted for code generation in order to be able to use the generated annotation processors independently of {\em Ann} and its tooling. Moreover, the generated processors would normally be more efficient than processors based on interpretation of {\em Ann} models.

The next three subsections describe the abstract, concrete syntax and semantics of the Ann DSL. 


\subsection{Abstract syntax}

\begin{figure}[h!]
\centering
	\includegraphics[width = 8cm]{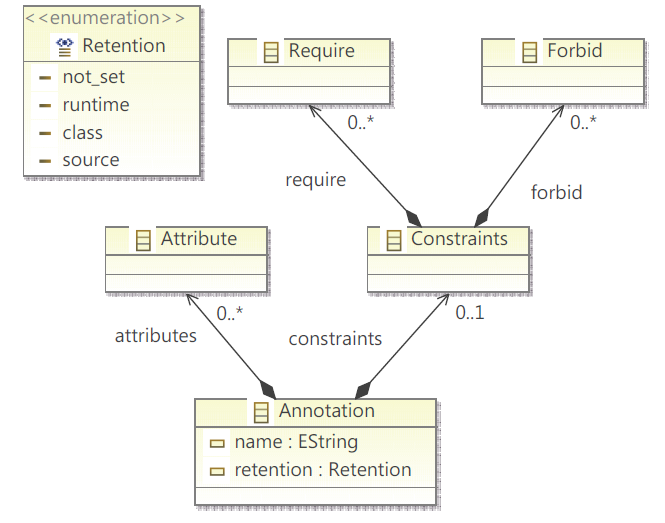}
  \caption{Simplified meta-model excerpt representing the abstract syntax of Ann.}
	\label{metam}
\end{figure}

The simplified meta-model that describes the abstract syntax of Ann can be found in 
	Figure \ref{metam}. 

The {\tt Annotation} meta-class contains both the attributes of an annotation and its associated constraints. Note that an annotation with no constraints is allowed, as in the Java language.

\begin{figure}[h!]
\centering
	\includegraphics[width = 13cm]{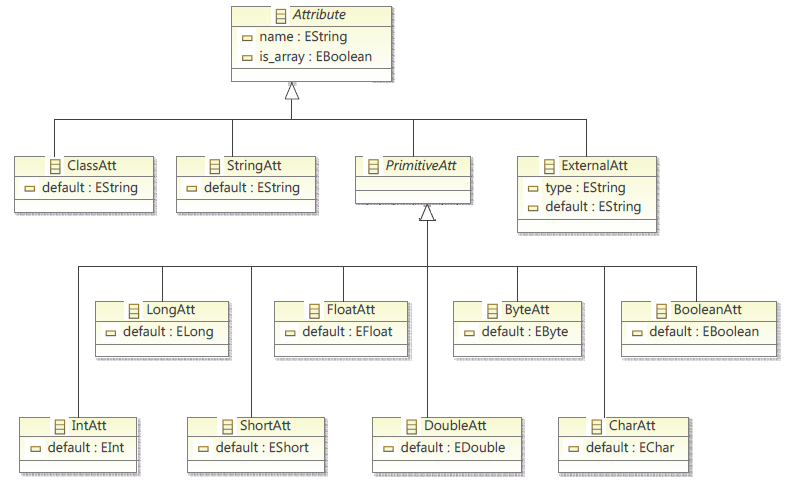}
  \caption{Meta-model excerpt for attributes.}
	\label{metam_attrs}
\end{figure}

Details concerning attributes are shown in Figure~\ref{metam_attrs}. Meta-class {\tt ExternalAttr} represents attributes declared
externally to Ann, including enumerated types and other annotations. We also consider all possible primitive types
for attributes, the possibility of default values and arrays. 

\begin{figure}[h!]
\centering
	\includegraphics[width = 8cm]{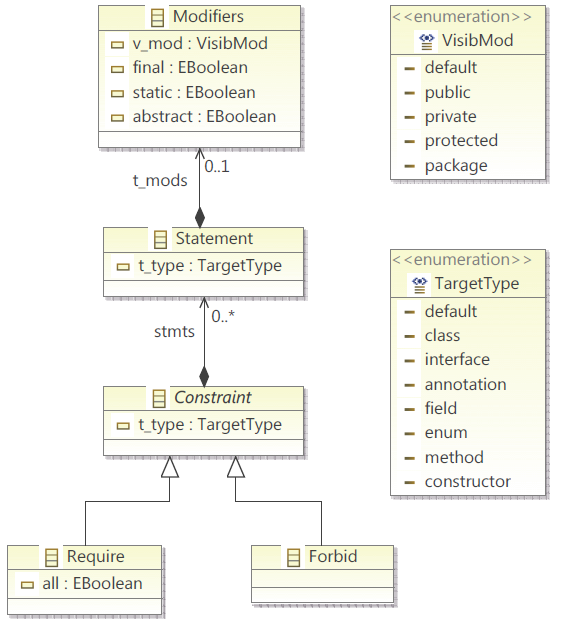}
  \caption{Meta-model excerpt for annotation constraints.}
	\label{metam-const}
\end{figure}

Constraints are split into two types: requirements (class {\tt Require} in Figure~\ref{metam}) and prohibitions (class {\tt Forbid} in Figure~\ref{metam}). Multiple constraints over the same annotation type have {\tt AND} semantics.
In Figure \ref{metam-const} we can see an expanded section of the meta-model of Figure~\ref{metam}, in particular the one concerning the constraints.
		
Each \emph{statement} represents a description of a Java element (like {\tt class}, {\tt interface} or {\tt field}) 
over which the annotation is (dis-)allowed. Ann supports the characterization of elements regarding their visibility, and
whether they should be {\tt final}, {\tt static} or {\tt abstract}.
In order to enhance the expressive power of Ann, several statements are possible within the same constraint (e.g., if the same annotation can be applied to several targets). 
This is why in the case of requirements multiple statements have {\tt OR} semantics; whereas {\tt AND} semantics are applied in the case of prohibitions. 
Note that {\tt AND} semantics for requirements would not add any additional expressive power since this is already granted by the multiplicity of {\tt Require} objects. 
In the case of prohibitions, the expressive power is enhanced by allowing to forbid \textit{simultaneous} characteristics in a Java target element: this kind of constraints are only violated if all the statements are satisfied by such element.

There is also the possibility of expressing constraints for specific target types (e.g., a {\tt field}), 
which indicates that the given constraint only applies when the annotation is attached to that target type (e.g., a {\tt field}). { This corresponds to the attribute {\tt t\_type} of {\tt Constraint}.} 
An annotation is { thus} correctly placed at a target type if it satisfies some of the statements of the  
requirements whose {\tt t\_type} coincides with the given target, and none of the respective prohibition statements. 

For these restricted types of constraints, there is a conceptual distinction depending on whether {\tt t\_type} is a Java container or inner type, since the statements will
refer to characteristics of its inner or containing elements, respectively. For example, if the target type is \texttt{field}, then the statements will constrain the classes, interfaces or annotations that contain it{; that is, the attribute {\tt t\_type} of an {\tt Statement} inside such {\tt Constraint} can only be one of {\tt class}, {\tt interface} or {\tt annotation}\footnote{This conceptual remark is also checked when validating the constraints.}.

These two types of constraints, and their combinations, provide enough expressive power to cover a large scope of the conceptual model behind a set of dependent annotations; as it will be shown in Section \ref{usecase}, in a real use case.

Figure~\ref{example-annot} shows an example annotation in abstract syntax. The annotation has {\tt Person} as name
and declares three attributes: {\tt name}, {\tt age} and {\tt weight}. It declares two constraints: one requiring a {\tt public class}
and another one forbidding the class to have {\tt final fields}. This is an example of an annotation placed at a 
container Java type (\texttt{class}) with statements that constrain its inner components (\texttt{field}s are forbidden to be \texttt{final}).

\begin{figure}[h!]
\centering
	\includegraphics[width = 12cm]{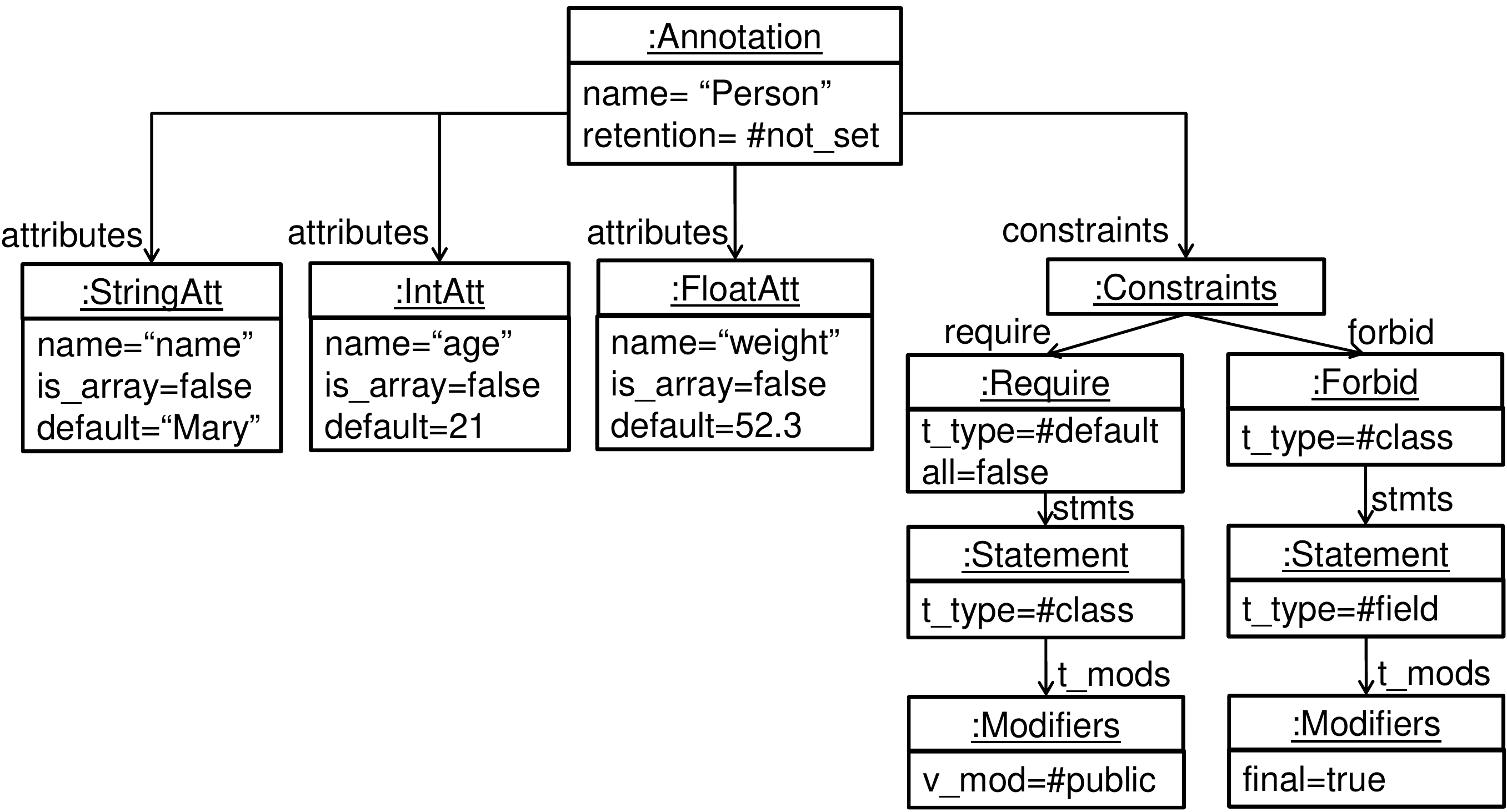}
  \caption{Example annotation definition in abstract syntax.}
	\label{example-annot}
\end{figure}

\subsection{Concrete syntax}


We have designed a textual concrete syntax for Ann. An excerpt of the concrete syntax definition for the constraints within an annotation can be found in Listing \ref{sint_c},
represented in Extended Backus-Naur Form (EBNF). Appendix \ref{an:sint} includes the full definition.

\begin{lstlisting}[
	style=EBNF, 
	language=EBNF, 
	caption=Concrete syntax excerpt for constraints in Ann.,
	label=sint_c
]
Forbid ::= "forbid" Statement ("and" Statement)* ";"; 
			| "at" TargetType : "forbid" Statement ("and" Statement)* ";";

Require ::= "require" Statement ("or" Statement)* ";";
			| "at" TargetType : "require" "all"? Statement ("or" Statement)* ";";
\end{lstlisting}

Listing \ref{ann_def} shows how the Java annotation type \texttt{Person} previously shown in Listing \ref{java_def}, and in
Figure~\ref{example-annot}, is described using the concrete syntax of Ann. 
A new keyword (\texttt{annotation}) is used on its declaration (line 3). Instead of
using methods to define the annotation parametrs (c.f. Listing~\ref{java_def}), we use the regular 
Java syntax for definining class attributes (lines 4--6 of Listing~\ref{ann_def}). 

\begin{lstlisting}[
	style=JavaColors, 
	language=Annotation, 
	caption=Annotation \texttt{Person} defined in Ann.,
	label=ann_def
]
package examples;

annotation Person {
    String name = "Mary";	
    int age = 21;
    float weight = 52.3;
		
    require public class;          // annotation allowed for classes...
		
    at class: forbid final field;  // ... with no final fields
}
\end{lstlisting}

Regarding the restriction of the allowed targets, we can now express some
more elaborated constraints, in this case that Person can only annotate public classes (line 8)
with no final fields (line 10).
We recall that with Java the closest we can get to this statement is that the annotation could have as targets
classes, interfaces and enumerations, which is less specific than what we obtain with this annotation type definition.


    %
		%

In the concrete syntax for requirements, we also note the special keyword {\tt all}. This would apply if, for instance, we would want
that all the methods of the classes annotated with {\tt Person} were also public. Then we would add the clause \code{at class: require all public method}.

\subsection{Semantics: code generation}

In order to fully specify the semantics of {\em Ann} it is necessary to generate on the one hand the Java code associated with the definition of
the annotations; and on the other hand the code of the processors. The latter will ensure that the constraints specified for 
each of the defined annotations are fulfilled.

For each of the annotations defined at most two processors will be generated, one for checking the requirements and the other for 
checking the prohibitions. The structure of the annotation processors generated complies with the one presented in Section \ref{javaann}:
each of the relevant elements of the Java program is looked up to check whether its properties satisfy the specified 
requirements or prohibitions.

\subsection{Tool support}\label{sec:emf}

The different components of the Ann DSL have been developed using the Eclipse Modelling Framework (EMF)~\cite{EMF}.

The meta-model has been described with the meta-modelling language Ecore\footnote{\url{http://www.eclipse.org/ecoretools/}}, which is based on a subset
of UML class diagrams for the description of structural aspects.
The Xtext\footnote{\url{http://www.eclipse.org/Xtext/}} framework, integrated with EMF 
and able to generate a fully customisable and complete editor for the defined language, has been used to define the textual concrete syntax. 
Finally, the code generator has been developed using Xtend\footnote{\url{https://eclipse.org/xtend/}}, a dialect of the Java language included in Xtext. Xtend is more expressive and flexible than Java and has facilities for model navigation. It also
allows creating generation templates, what makes it specially useful for code generation. 
The tool also integrates the USE validator, in order to check constraint conflicts. The use of such model finder will be explained in section~\ref{validation}.

\begin{figure}[h!]
\centering
	\includegraphics[width = 9cm]{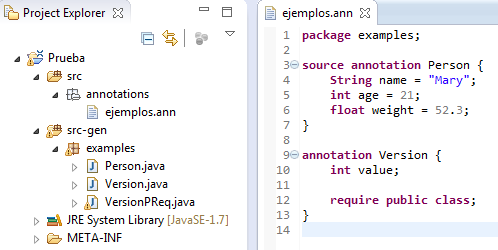}
  \caption{The Ann IDE.}
	\label{ann-IDE}
\end{figure}

The result is an Eclipse plug-in, which is seamlessly integrated within the Eclipse Java Development
Tools (JDT)\footnote{\url{http://www.eclipse.org/jdt/}}. A screenshot of the IDE is shown in Figure~\ref{ann-IDE}.

\section{Annotation validation}\label{validation}
The advantage of using a high-level language, like {\em Ann}, to make explicit the integrity constraints of a set
of annotations, is that they are amenable to analysis. We use model finding techniques for this purpose~\cite{Jackson,USE}.
Model finders are tools supporting a high-level notation to describe features of models, and use 
constraint solving to find a model exhibiting such features. Typically, model features are described 
using structural data models (e.g., class diagrams with OCL constraints~\cite{USE}, or 
relational logic~\cite{Jackson}), and rely on lower-level SAT or SMT solver engines (like 
KodKod~\cite{Kodkod} or Z3~\cite{Z3}). Solvers typically perform a bounded search, so that 
only models up to a given size are sought. Nonetheless, according to the ``small scope'' 
hypothesis~\cite{Jackson}, a large proportion of errors in a system can be identified by 
considering only instances within a small scope. We use the USE model finder, which accepts
as input a meta-model plus OCL invariants.

In order to perform the analysis, we have created a simplified meta-model of Java, containing only the elements that we consider in the Ann language. Then, annotation constraints are translated into OCL, and USE is employed to search for an instance of the previous Java meta-model satisfying all constraints. If no such model is found, then the annotation constraints are incompatible. Moreover,
we can also search for Java models containing a combination of annotations of interest. This can be done
by another OCL constraint explicitly demanding the occurrence of the desired combination of annotations.

\begin{figure}[t!]
\centering
	\includegraphics[width = 12cm]{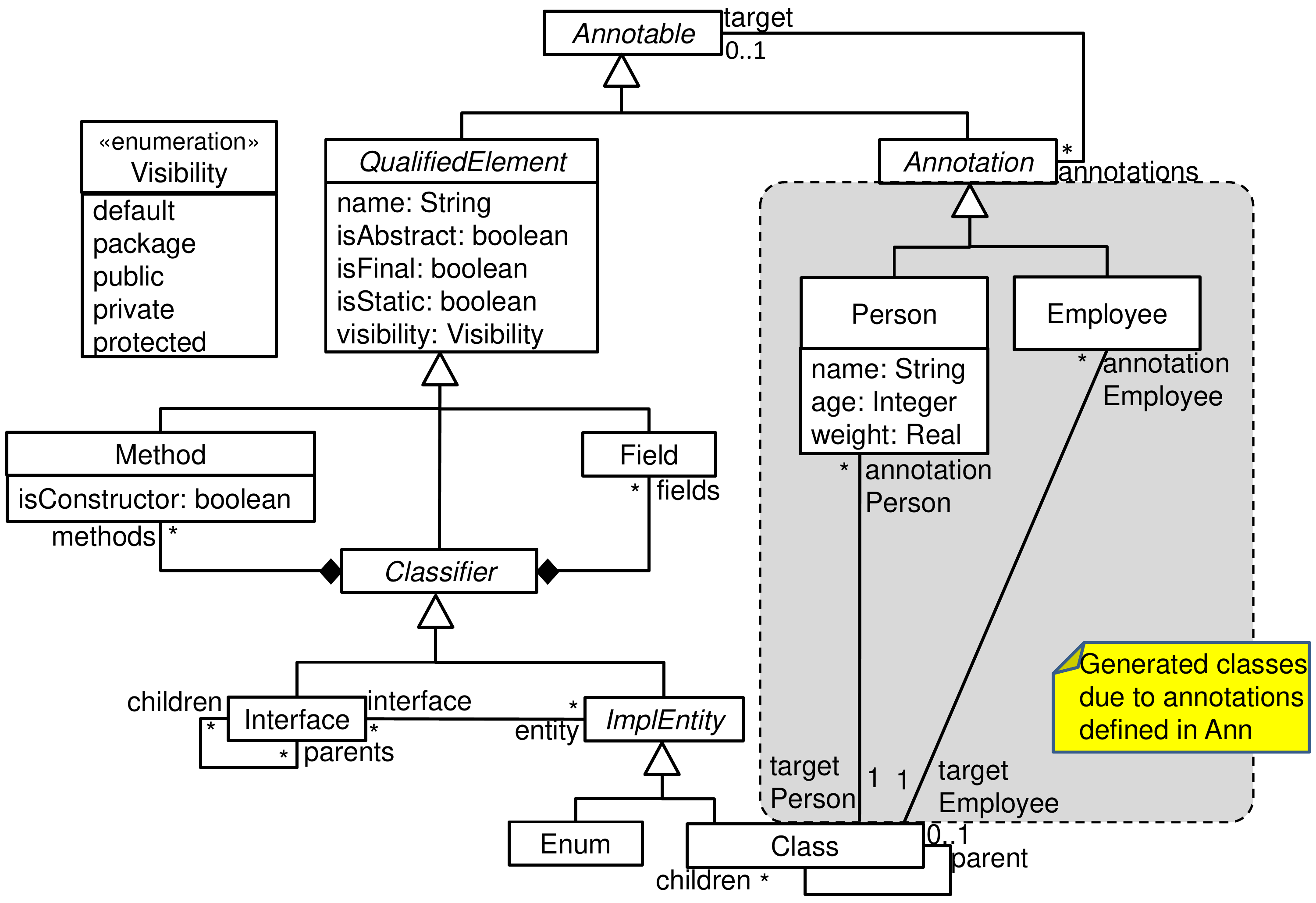}
  \caption{Java meta-model used for annotation validation and verification. Encircled in a dotted region, it contains two classes generated from the annotations \texttt{Person} and \texttt{Employee}.}
	\label{jmm}
\end{figure}

Figure~\ref{jmm} shows the Java meta-model we use for the validation and verification of annotations. 
A few OCL constraints have been added to different classes, for
example restricting the visibility of {\em Class} to be {\em default} or {\em public}; demanding {\em abstract} methods to reside in {\em abstract} {\em Class}es; and forbidding cycles of class and interface inheritance.

For the analysis, the main idea is to enrich such meta-model with classes and constraints generated from the annotation definitions. If the resulting meta-model is satisfiable, then there is no conflict in the annotation definition. For example, Figure~\ref{jmm} shows two classes: \texttt{Person} and \texttt{Employee} generated from two annotation definitions. The first one from the definition in Listing~\ref{ann_def}, while the second one just requires classes with package visibility and with a {\em Person} annotation. In Figure~\ref{jmm}, these generated classes are shown encircled in a dotted region. For class \texttt{Person} we generate an association to \texttt{Class}, which is its only allowed target. If an instance of \texttt{Person} is created, it needs to annotate a \texttt{Class} and therefore the cardinality of role \texttt{targetPerson} is \texttt{1}. Similarly, another association is created for \texttt{Employee}. 
Additionally, the OCL constraints of Listing~\ref{OCL_Gen} are generated for both classes.

\begin{figure}[t!]
\begin{lstlisting}[
      style=JavaColors,
	language=OCL, 
	caption=Generated OCL constraints from the \texttt{Person} and \texttt{Employee} annotations,
	label=OCL_Gen
]
-- Generated invariants for Person annotation
context Person 
	inv redefs: -- redefines Annotation.target
   		self.target.isUndefined()
context Person 
	inv require_public_class: -- requires public classes
   		self.targetPerson.visibility=#public
   -- at annotated classes, forbids final fields
context Person 
	inv at_class_forbid_final_field: 
   		self.targetPerson.fields->forAll(a | a.isFinal = false )

-- Generated invariants for Employee annotation
context Employee 
	inv redefs: -- redefines Annotation.target
   		self.target.isUndefined()
context Employee 
	inv require_annPerson_package_class: -- requires package visibility
   		self.targetEmployee.visibility=#package and 
   		self.targetEmployee.annotationPerson->notEmpty()
\end{lstlisting}
\end{figure}

The two invariant named \texttt{redefs} in lines 2-4 and 14-16 emulate the redefinition of the \texttt{target} role by \texttt{targetPerson} and \texttt{targetEmployee} in classes \texttt{Person} and \texttt{Employee} respectively. 
This is necessary, because for both annotations, \texttt{Class} is their only allowed target. { Note that the \texttt{target} role
is useful for annotations that do not explicitly declare a target, so that they can be placed anywhere}.
For \texttt{Person} we require the target class to be public (invariant \texttt{require\_public\_class} in lines 5-7), and to have no final fields (invariant \texttt{at\_class\_forbid\_final\_field} in lines 9-11). For \texttt{Employee} 
we require the target class to have \texttt{package} visibility (invariant \texttt{require\_annPerson\_package\_class}, line 19), and be also annotated with the \texttt{Person} annotation (invariant \texttt{require\_} \texttt{annPerson\_package\_class}, line 20).

Feeding the meta-model of Figure~\ref{jmm} and the generated constraints in Listing~\ref{OCL_Gen} to USE, it returns no model, and hence the constraints are unsatisfiable. On reflection, we realize that the designed constraints for the annotations are in conflict, because annotation \texttt{Employee} demands classes with \texttt{package} visibility, and to be annotated with the \texttt{Person} annotation, which requires classes with \texttt{public} visibility. Designers can then modify the constraints, for example requiring \texttt{Employee} to annotate \texttt{public} classes. Alternatively, they might drop the constraint on visibility, because it would be redundant with the similar constraint from \texttt{Person}. 
{ While this example is simple, in more complex cases, the user has the burden to find the reasons for the conflict. This is typically done 
by systematically trying all combinations of constraints within the considered set (by manually disabling combinations of constraints). Automated, more efficient support for this task
(e.g., like the method proposed in~\cite{PrzigodaWD15}) is left for future work.
Similarly, constraint redundancy can be (manually) investigated with solvers like USE~\cite{GogollaHK10}, but automating this task is also left for future work.}

Once the constraint for \texttt{package} visibility is deleted from \texttt{Employee}, USE would return a model like the one in Figure~\ref{USE_model}(a), proving that there is no conflict, and provide the designer with an example of use of the designed annotation. Figure \ref{USE_model}(b) shows a representation of the USE model in the textual syntax of Java, which could be useful to the designer, to see an example of use of the designed annotations.

\begin{figure}[t!]
\centering
	\includegraphics[width = 10cm]{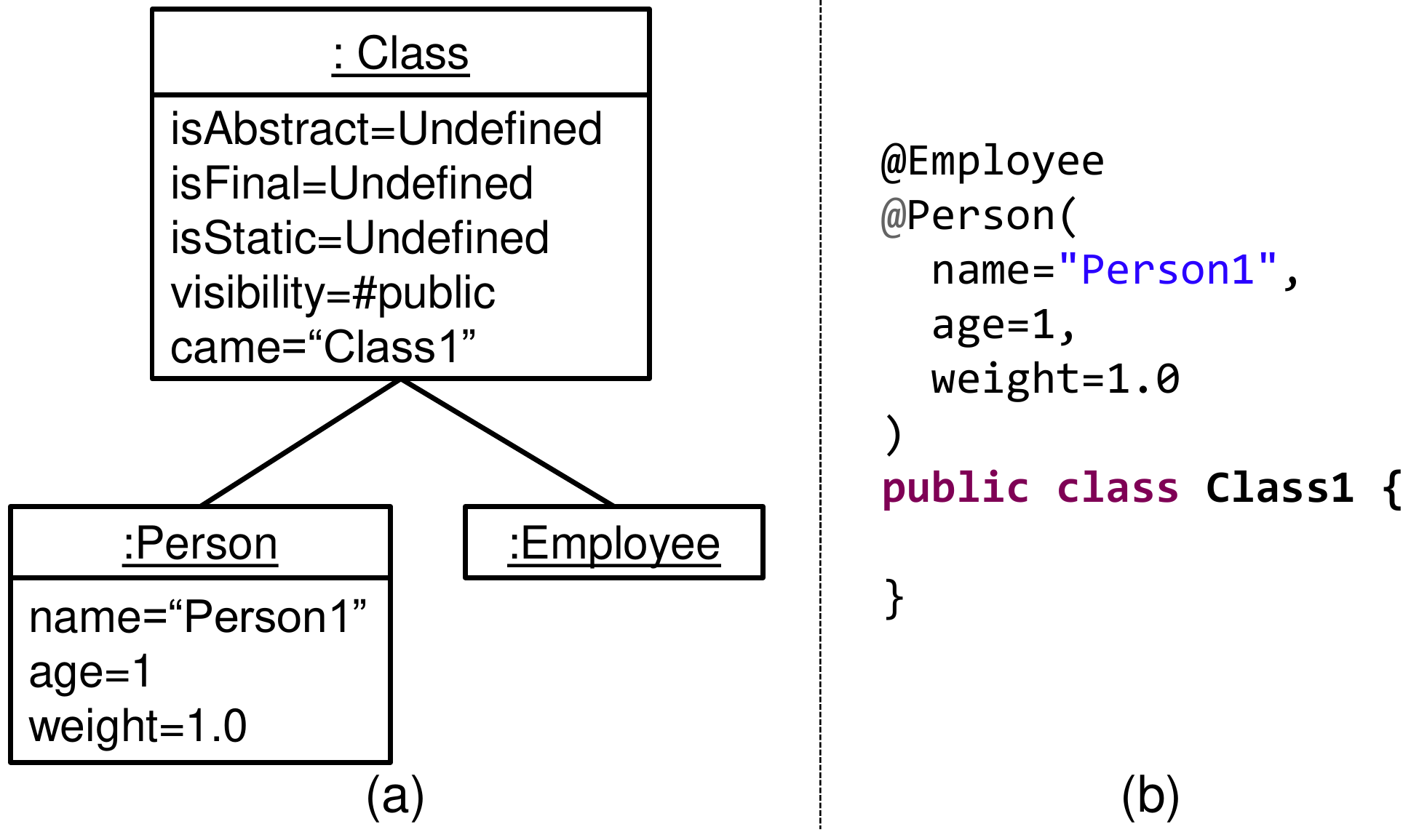}
  \caption{(a) Model produced by USE. (b) Java serialization of the model.}
	\label{USE_model}
\end{figure}

We next provide a systematic description of how the Java meta-model is to be extended and how the OCL invariants are generated given an Ann model. 
For each Ann annotation named \texttt{$\langle$ANN$\rangle$}, we create a subclass
of \texttt{Annotation} named \texttt{$\langle$ANN$\rangle$}. 
Then, additional OCL code is generated for the annotation's \texttt{require} and \texttt{forbid} constraints.

\subsection{Generation of target requirements}
The code generation scheme for \texttt{require} constraints is shown in Table~\ref{tbl:ann}. 
The second column of the table shows the generated OCL, as well as the extra elements to be
created in the meta-model (using the textual notation of USE for class diagrams).

\noindent
\begin{table}[t!]
\begin{centering}
\scriptsize
\begin{tabular}{| p{4.6cm} | p{8.2cm} |}
\hline
{\bf Ann} (scheme) 				& {\bf New meta-model elements and OCL} \\ \hline
\begin{lstlisting}[language=Annotation, style=intable]
annotation <ANN> {
  require 
    <MODS1> <TYPE1>
  ...
  require 
    <MODSn> <TYPEn>
}
\end{lstlisting} 
&
\begin{lstlisting}[language=OCLUSE, style=intable]
-- Assoc. created for each distinct type
association <ANN>target<ANN><TYPEi> between
  <TYPEi> [0..1] role target<ANN><TYPEi>
  <ANN> [0..*] role annotations<ANN>
end

context <ANN> inv redefs: 
  self.target.isUndefined()

-- if some modifier is not empty
context <ANN> inv <inv_name>: 
  self.target<ANN><TYPEi>->notEmpty() 
    implies (
      -- and-catenation of the expressions  
      -- of the subcases below
    )
\end{lstlisting}

\\ \hline
\multicolumn{2}{|l|}{{\bf Subcases:}} 
\\ \hline
\begin{lstlisting}[language=Annotation, style=intable]
if (public or package or
   private or protected) 
  in <MODSi>
\end{lstlisting} 
&
\begin{lstlisting}[language=OCLUSE, style=intable]
   self.target<ANN><TYPEi>.visibility = 
     # <public|package|private|protected>
\end{lstlisting}
\\ \hline
\begin{lstlisting}[language=Annotation, style=intable]
if abstract 
  in <MODSi>
\end{lstlisting} 
&
\begin{lstlisting}[language=OCLUSE, style=intable]
  self.target<ANN><TYPEi>.isAbstract 
\end{lstlisting}
\\ \hline
\begin{lstlisting}[language=Annotation, style=intable]
if static 
  in <MODSi>
\end{lstlisting} 
&
\begin{lstlisting}[language=OCLUSE, style=intable]
  self.target<ANN><TYPEi>.isStatic 
\end{lstlisting}
\\ \hline
\begin{lstlisting}[language=Annotation, style=intable]
if final 
  in <MODSi>
\end{lstlisting} 
&
\begin{lstlisting}[language=OCLUSE, style=intable]
  self.target<ANN><TYPEi>.isFinal
\end{lstlisting}
\\ \hline

\end{tabular}
\caption{Translating basic Ann annotation constraints (requirements) into OCL.}\label{tbl:ann}
\end{centering}
\end{table}

%
%

As explained in Section \ref{dslann}, requirements have an \texttt{OR} semantics. This means that some should be satisfied by the annotation. This way,
we create an association between the generated class \texttt{$\langle$ANN$\rangle$} and every distinct \texttt{$\langle$TYPEi$\rangle$}.
In the case of more than one requirement, we set the role cardinality in the part of the type to $0..1$. If there is only
one requirement, we set it to $1$, as shown in Figure~\ref{jmm} (and the name of the role is also simplified to \texttt{target$\langle$TYPEi$\rangle$}).
Then, we conceptually redefine the generic \texttt{target} association end to a particular one
by means of invariant \texttt{redefs}.

In case some \texttt{require} clause has a modifier, we generate another invariant for the class \texttt{$\langle$ANN$\rangle$}.
The table shows the different possible modifiers, regarding visibility, or demanding the type to be abstract, static or final. All the stated modifiers apply to the \texttt{$\langle$TYPEi$\rangle$}, and therefore 
we and-concatenate the sub-expressions generated by
every non-empty modifier. This is done by first checking if the corresponding association role is not empty.
This checking is not needed if there is only one requirement (as in line 5 of Listing~\ref{OCL_Gen}).
Note that lines 11-16 in Table \ref{tbl:ann} represent the OCL invariant that corresponds to one single statement. In the event of more than one statement in a \texttt{require} (as the Ann syntax allows), the generated OCL would correspond to the or-concatenation of those lines, for each statement. For future references, we will denote those lines as \texttt{condition($\langle$TYPEi$\rangle$, $\langle$MODSi$\rangle$)}.


\subsection{Generation of co-occurrence constraints for annotations}

\noindent
\begin{table}[t!]
\scriptsize
\begin{tabular}{| p{4.5cm} | p{8.5cm} |}
\hline
\multicolumn{2}{|l|}{{\bf Subcases (co-occurrence requirements):}}  \\ \hline
\begin{lstlisting}[language=Annotation, style=intable]
annotation <ANN> {
  ...
  require <ANN1> 
  ...
}
\end{lstlisting} 
&
\begin{lstlisting}[language=OCLUSE, style=intable]
-- Invariant created for each distinct type
context <ANN> inv <inv_name>: 
 self.target<ANN><TYPEi>->notEmpty() implies
 self.target<ANN><TYPEi>
   .annotations<ANN1>->notEmpty()
\end{lstlisting}

\\ \hline

\begin{lstlisting}[language=Annotation, style=intable]
annotation <ANN> {
  ...
  require <ANN1> 
    <MODS1> 
    <TYPE>
  ...
}
\end{lstlisting} 
&
\begin{lstlisting}[language=OCLUSE, style=intable]
context <ANN> inv <inv_name>: 
 self.target<ANN><TYPE>->notEmpty() implies
 (self.target<ANN><TYPE>
   .annotations<ANN1>->notEmpty()
  and condition(<TYPE>, <MODS1>))
\end{lstlisting}
\\ \hline
\begin{lstlisting}[language=Annotation, style=intable]
// <TYPE1> is contained  
// in <TYPE>
annotation <ANN> {
  ...
  at <TYPE> : 
  require <ANN1> 
    <MODS1> 
    <TYPE1>
  ...
}
\end{lstlisting} 
&
\begin{lstlisting}[language=OCLUSE, style=intable]
context <ANN> inv <inv_name>: 
self.target<ANN><TYPE>.notEmpty() implies
 <ANN1>.allInstances()->exists( 
  t | t.target<ANN1><TYPE1>.owner = 
   self.target<ANN><TYPE>
  and condition(<TYPE1>, <MODS1>))
\end{lstlisting}
\\ \hline
\begin{lstlisting}[language=Annotation, style=intable]
// <TYPE> is contained  
// in <TYPE1>
annotation <ANN> {
  ...
  at <TYPE> : 
  require <ANN1> 
    <MODS1> 
    <TYPE1>
  ...
}
\end{lstlisting} 
&
\begin{lstlisting}[language=OCLUSE, style=intable]
context <ANN> inv <inv_name>: 
 self.target<ANN><TYPE>.notEmpty() implies
 <ANN1>.allInstances()->exists( 
  t | t.target<ANN1><TYPE1> = 
   self.target<ANN><TYPE>.owner 
  and condition(<TYPE1>, <MODS1>))
\end{lstlisting}
\\ \hline
\end{tabular}
\caption{Translating Ann co-occurrence constraints for annotations (requirements) into OCL.}\label{tbl:ann2}
\end{table}

Table~\ref{tbl:ann2} shows the scheme of the invariants generated for requirements of co-occurrence of annotations. 
In the first case, annotation \texttt{$\langle$ANN$\rangle$} requires the occurrence of 
\texttt{$\langle$ANN1$\rangle$} in every place where \texttt{$\langle$ANN$\rangle$} may appear. Hence, we check
that every valid target of \texttt{$\langle$ANN$\rangle$} is also annotated with \texttt{$\langle$ANN1$\rangle$}.

In the second case, annotation \texttt{$\langle$ANN1$\rangle$} acts as a constraint on a type where the annotation \texttt{$\langle$ANN1$\rangle$} can be placed. This case is direct as it is analogous to the ones we
explained for modifiers in Table \ref{tbl:ann}. 

The third and fourth cases deal with the situation 
when at a certain target, one of its contained or containing elements (see Section \ref{dslann} for an explanation of this distinction) should
be annotated with \texttt{$\langle$ANN1$\rangle$}, respectively. Specifically, we check that whenever \texttt{$\langle$ANN$\rangle$} is annotating target \texttt{$\langle$TYPE$\rangle$}, there is an occurrence of target \texttt{$\langle$TYPE1$\rangle$} with the specified constraints, that is, with \texttt{$\langle$MODS1$\rangle$} and annotated with \texttt{$\langle$ANN1$\rangle$}. The fourth case is the converse.

Again, if there is only one possible target type, then
there is no need to check the role that is not empty (because the role would have \texttt{1..1} cardinality). In our example in Listing~\ref{OCL_Gen}, there is no need to test
that the role \texttt{targetEmployee} is not empty in line 20. Finally, recall from Section \ref{dslann} that in presence of multiple statements in the same
\texttt{require}, the semantics is to or-concatenate them within the same invariant.

\subsection{Generation of requirements for targets}

Table~\ref{tbl:ann3} describes the OCL equivalent to ``at'' constraints (for the \texttt{require} case). These kinds of constraints describe structural or positional (depending on whether the target is a container or contained type, respectively) requirements or prohibitions for the targets of a given annotation (see Section \ref{dslann} for more information).

The scheme of translation is
similar to the one of Table~\ref{tbl:ann2}, specially the two last cases, so for simplicity we only consider the case of a container type (the case for \texttt{method} and \texttt{constructor}, because the
case of \texttt{field} is analogous), and the use of the \texttt{all} modifier. We will not consider the appearance of annotations in the 
statements because that case has already been treated in Table \ref{tbl:ann2}.

\noindent
\begin{table}[t!]
\scriptsize
\hskip-0.5cm
\begin{tabular}{| p{3.5cm} | p{10.5cm} |}
\hline
\multicolumn{2}{|l|}{{\bf ``at'' requirements for method and constructor targets}}  \\ \hline
\begin{lstlisting}[language=Annotation, style=intable]
annotation <ANN> {
  ...
  at <TYPE1>: 
    require all 
      <MOD> method
  ...
}
\end{lstlisting} 
&
\begin{lstlisting}[language=OCLUSE, style=intable]
context <ANN> inv <inv_name>: 
   self.target<ANN><TYPE1>->notEmpty() implies
     self.target<ANN><TYPE1>.methods->forAll( m |
       condition(Method, <MOD>))
\end{lstlisting}

\\ \hline
\begin{lstlisting}[language=Annotation, style=intable]
annotation <ANN> {
  ...
  at <TYPE1>: 
    require all 
      <MOD> 
      constructor
  ...
}
\end{lstlisting} 
&
\begin{lstlisting}[language=OCLUSE, style=intable]
context <ANN> inv <inv_name>: 
   self.target<ANN><TYPE1>->notEmpty() implies
     self.target<ANN><TYPE1>.methods->forAll( m |
       m.isConstructor implies condition(Method, <MOD>))
\end{lstlisting}

\\ \hline

\begin{lstlisting}[language=Annotation, style=intable]
annotation <ANN> {
  ...
  at <TYPE1>: 
    require 
     <MOD> method
  ...
}
\end{lstlisting} 
&
\begin{lstlisting}[language=OCLUSE, style=intable]
context <ANN> inv <inv_name>: 
   self.target<ANN><TYPE1>->notEmpty() implies
     self.target<ANN><TYPE1>.methods->exists( m |
       condition(Method, <MOD>))
\end{lstlisting}

\\ \hline

\begin{lstlisting}[language=Annotation, style=intable]
annotation <ANN> {
  ...
  at <TYPE1>: 
    require 
     <MOD> 
     constructor
  ...
}
\end{lstlisting} 
&
\begin{lstlisting}[language=OCLUSE, style=intable]
context <ANN> inv <inv_name>: 
   self.target<ANN><TYPE1>->notEmpty() implies
     self.target<ANN><TYPE1>.methods->exists( m |
       m.isConstructor and condition(Method, <MOD>))
\end{lstlisting}

\\ \hline

\end{tabular}
\caption{Translating Ann ``at'' constraints (requirements) into OCL.}\label{tbl:ann3}
\end{table}

We have presented only the translation for \texttt{require} constraints, because the case for \texttt{forbid} are analogous, but it differs by using negation, and conjunction instead of disjunction when in presence of several statements.

\newpage


\section{Evaluation}\label{usecase}
In this section we evaluate two aspects of our approach. On the one hand the expressivity, usefulness and advantages of Ann by modelling a subset of the JPA annotations.
On the other, we provide an evaluation of the performance and scalability of the constraint-based validation of the annotations. 
Finally, we discuss possible threats to validity with respect to the evaluation performed.

\subsection{A real use case: JPA annotations}

In order to test the Ann DSL, we have chosen a subset of the JPA annotations, namelly
{\tt Entity}, {\tt Id}, {\tt IdClass}, {\tt Embeddable} and {\tt EmbeddedId}. 
This selection has been made according to their extensive use in the JPA context, given that all of them
are used to describe entities and their primary keys, central concepts in database design.

\subsubsection{Characteristics of the set of annotations}
The characteristics that the targets of the {\tt Entity} annotation must comply with were outlined in Section \ref{javaann}. 
Given that this annotation defines an entity within a database, a corresponding primary key must also be specified within those targets. The other
selected annotations are used precisely for this purpose.

There are two alternatives in JPA for representing compound primary keys. Both of them involve using a class
which contains the fields that compose the primary key.
In the first approach, the class can be annotated with \texttt{Embeddable}, in which case it represents
a class whose instances are intrinsic components of the original entity. They share with it the primary key and
the class is used as a field on the entity it is embedded, annotated with \texttt{EmbeddedId}, which marks it as
primary key. Figure \ref{val:jpaembid} shows an example of this alternative.

\begin{lstlisting}[style=JavaColors, language=Java, caption=Primary key with \texttt{EmbeddedId}., label=val:jpaembid]
@Embeddable
public class EmployeePK implements Serializable {
    private String name;
    private long id;

    public EmployeePK() {
    }
    ...
}

@Entity
public class Employee implements Serializable {
    @EmbeddedId EmployeePK primaryKey;
 
    public Employee() {
    }
    ...
}
\end{lstlisting}

The other approach is to use a 
class that represents the fields of the primary key, but is not embedded. In this case, when defining the entity we use the \texttt{IdClass} annotation to indicate the class that contains the fields of the compound primary key. Each of those fields is then added to the entity by using the \texttt{Id} annotation, as standard. Listing \ref{val:jpaidclass}
illustrates this other alternative with an example.

\begin{lstlisting}[style=JavaColors, language=Java, caption=Primary key with \texttt{IdClass}., label=val:jpaidclass]
public class EmployeePK implements Serializable {
    private String name;
    private long id;

    public EmployeePK() {
    }
    ...
}

@IdClass(EmployeePK.class)
@Entity
public class Employee {
    @Id String name;
    @Id long id; 
...
}
\end{lstlisting}

It is important to notice that a class cannot have a field or method annotated with \texttt{Id} and another one annotated with \texttt{EmbeddedId},
since that would imply two primary keys.

\subsubsection{Defining the annotations with Ann}
 
Listing \ref{jpa_cod} shows the description of the explained annotations using the Ann DSL.

	\begin{lstlisting}[style=JavaColors, language=Annotation, caption=Selected JPA annotations defined in Ann., label=jpa_cod]
runtime annotation Entity {
	String name = "";

	require class;
	forbid final class;
	
	at class: require public constructor or protected constructor;
	at class: forbid final method;
	
	at class: require @Id method or @Id field or 
                      @EmbeddedId method or @EmbeddedId field;
	at class: forbid @Id method and @EmbeddedId method;
	at class: forbid @Id field and @EmbeddedId field;
}

runtime annotation Embeddable {
	require class;
	
	at class: forbid @Id method;
	at class: forbid @EmbeddedId method;
	
	at class: forbid @Id field;
	at class: forbid @EmbeddedId field;
}

runtime annotation EmbeddedId {
	require method or field;
	
	at field: require @Entity class;
	at method: require @Entity class;
}

runtime annotation Id {
	require method or field;
	
	at field: require @Entity class;
	at method: require @Entity class;
}

runtime annotation IdClass {
	Class value;
	
	require @Entity class;
}
	\end{lstlisting}

Clearly 
the chosen subset of annotations is very interrelated given all the respective constraints that can be noticed. 
For example, a class annotated with {\tt Embeddable} (lines 16-24 in Listing~\ref{jpa_cod}) acts as a primary key for another class, in which it is embedded, 
and thus it must not have a primary key itself, prohibition which is expressed through lines 19-20 and 22-23 of Listing~\ref{jpa_cod}. 

Alternatively, the annotation {\tt IdClass} (lines 40-44 of Listing~\ref{jpa_cod}) can be used to specify the class that contains 
the fields which form the compound primary key. Therefore it can only be attached to classes annotated with {\tt Entity},
requirement described in line 43 of the listing. 

Annotations {\tt Id} (lines 33-38 of Listing~\ref{jpa_cod}) and {\tt EmbeddedId} (lines 26-31 of the same listing) mark the primary key of an entity, and thus can only
annotate methods or fields (lines 34 and 27 resp.) which form part of a class annotated with {\tt Entity} (lines 36-37 and
29-30 resp.).

Finally, regarding the {\tt Entity} annotation (lines 1-14 of Listing~\ref{jpa_cod}), structural properties of the annotated classes are expressed throughout
lines 4-8
; and lines 10-11 establish the need
of a primary key through a requirement, among other constraints.

After the definition of all the annotations and their constraints, we can check for inconsistencies using the constraint solver.
The generated OCL code is shown in Appendix~\ref{an:ocl}. In this case, USE reported no incompatibilities.
After this evaluation, the corresponding code can be generated and is ready to
use in both new or existing Java projects, as we will see in next subsection.

\subsubsection{Using the generated code}
The generated processors are capable of detecting where a constraint is being violated and also notify the developer by means
of an explanatory message. 

In Figure \ref{err1} the annotation {\tt Entity} is being used on a class and no primary key is being specified, which is a situation not allowed in the JPA context.

\begin{figure}
\centering
	\includegraphics[width = 12cm]{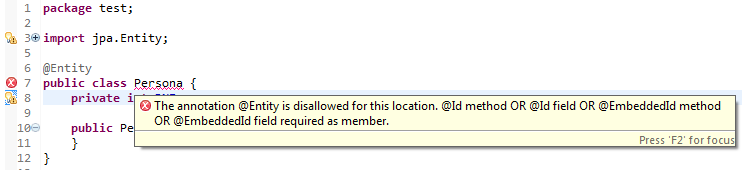}
  \caption{Entity without primary key.}
	\label{err1}
\end{figure}

Another example of misuse is the one shown in Figure \ref{err2}. In this case, the annotation {\tt Id} is used in a field inside 
a class that is not annotated as {\tt Entity}, which is a situation that leads to another error. This is analogous to Figure \ref{val:errclassent},
where this time the annotation \texttt{IdClass} is annotating a class which is not an entity.

\begin{figure}
\centering
	\includegraphics[width = 12cm]{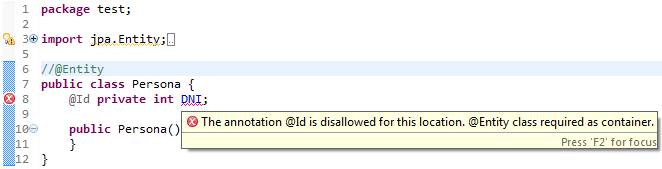}
  \caption{Primary key in a field not belonging to an entity.}
	\label{err2}
\end{figure}

\begin{figure}[t]
	\centering
		\includegraphics[width=13cm]{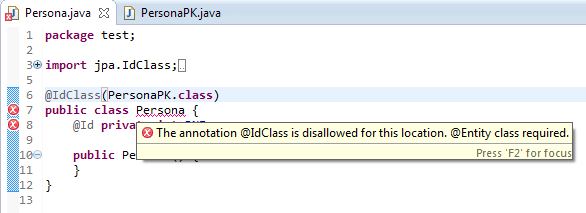}
  \caption{Using \texttt{IdClass} in a class which is not an entity.}
	\label{val:errclassent}
\end{figure}

In conclusion, we can see that, if we were to build this annotation set manually, we would have needed to:
(i) create the interfaces for the annotation in Java,
(ii) manually encode the integrity constraints of the annotation set in a Java annotation processor, and (iii) manually test 
the processor to find errors. By using Ann, steps (i) and (ii) were made with the DSL, and so there was no
need to manually program the annotation processor. To provide an intuition of the effort saved, the
generated processors amounted to 638 LOC.
The analysis of the annotation constraints was also made in an automated way, which avoided long, tedious cycles
of installing, testing and fixing the annotation processor.

\subsection{Efficiency and scalability of annotation validation}

 Constraint solving may generally involve costly computations, therefore we conducted an experiment to
	evaluate the efficiency of annotation validation to check the feasibility of its use in practice, and its scalability.
	
	For this purpose, we designed annotation sets of increasing size (from 2 to 64), and added constraints in each
	of them ranging from [1..2] constraints per annotation to [2..8] constraints per annotation. Constraints included
	both \texttt{require} and \texttt{forbid} constraints, and the former included co-ocurrence constraints involving all annotations within the set. Additionally,
	we designed both satisfiable and unsatisfible constraints. In the latter case, we opted for the ``difficult'' case,
	in which unsatisfiability is caused by a conflict between just two constraints. For the configuration of the search space, we used a bound of 1 to 4 instances
	of each annotation type.
	
	The experiments were performed on an Intel Core i7-2600 (3.4GHz) computer with 12GB RAM, and the version of USE was 4.1.1.
	Each experiment was repeated 5 times and we took the average time. 
	Figure~\ref{exp-graphic} shows a graphic summarizing the results, where the
	vertical axis shows the solving time in milliseconds and in logarithmic scale. The horizontal axis depicts the increasing number
	of annotations (2, 4, 8, 16, 32, and 64), with six series in each annotation set. The first two series contain the results of
	annotation sets with up to 2 constraints per annotation. SAT-2 corresponds to an annotation set with satisfiable constraints (a correct design),
	while UNSAT-2 is an annotation set with unsatisfiable constraints (an incorrect design). SAT-4 and UNSAT-4 depict
	annotation sets with half the annotations having 2 constraints and the other half 4 constraints. Similarly SAT-8 and UNSAT-8
	are sets with half the annotations having 2 constraints and the other half 8 constraints.
	The same results are shown in Table~\ref{tbl:solving-times}.

\begin{figure}[t!]
	\centering
	\includegraphics[width = 14cm]{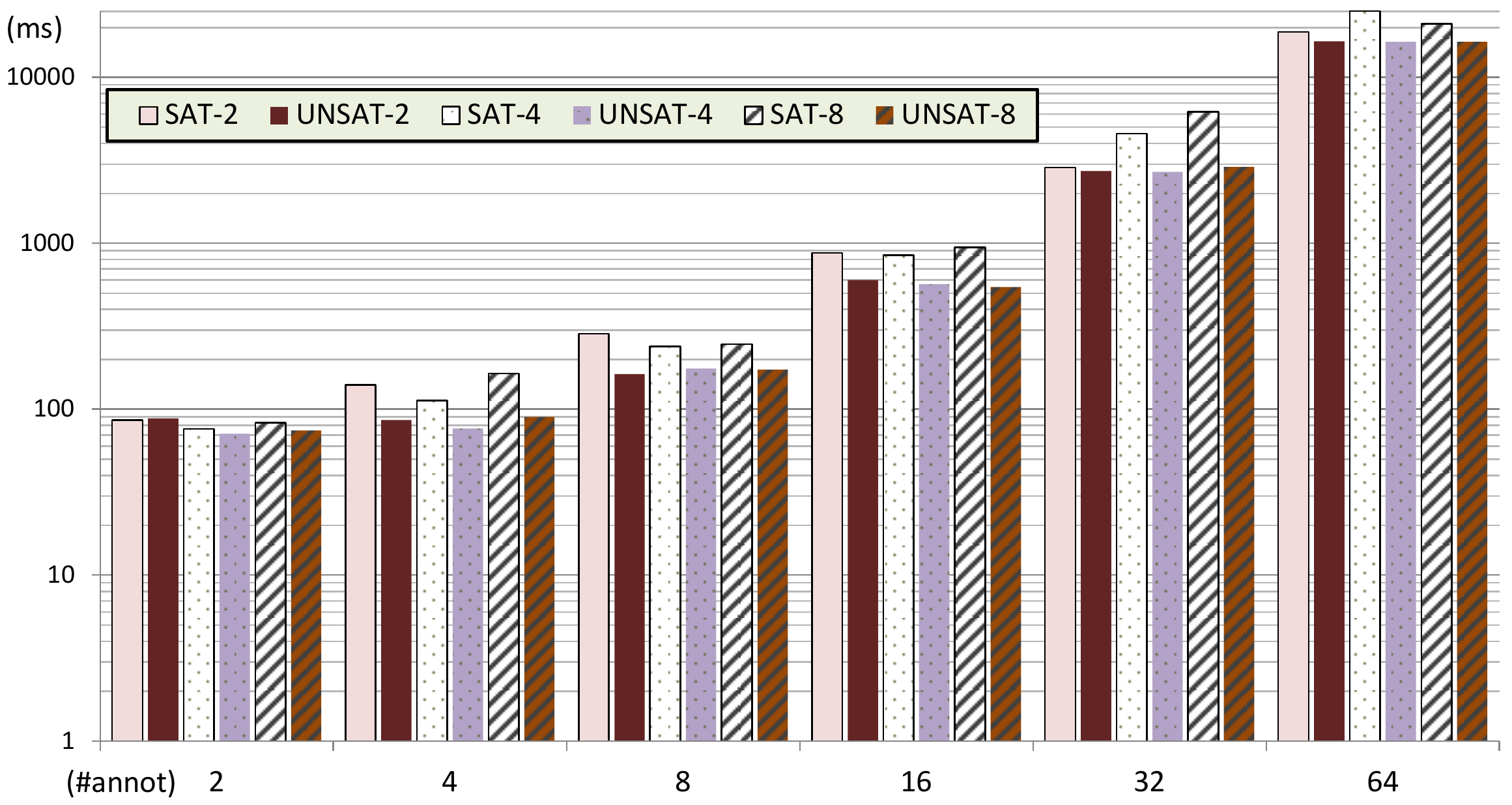}
	\caption{Solving times for increasing number of annotations and constraints within them.}
	\label{exp-graphic}
\end{figure}

\begin{table}[t!]
	\begin{center}
		\scriptsize
		\begin{tabular}{| p{2.5cm} | p{1.2cm} |  p{1.2cm} |  p{1.2cm} |  p{1.2cm} |  p{1.2cm} |  p{1.2cm} |}
			\hline
			& \multicolumn{6}{|c|}{{\bf \# annotations}} \\ \hline
			{\bf \# constraints} & {\bf 2} & {\bf 4} & {\bf 8} & {\bf 16} & {\bf 32} & {\bf 64} \\ \hline
			{\bf SAT-2}     & 86 & 140 & 286 & 874,8 & 2850,6 & 18802,8 \\ \hline
			{\bf UNSAT-2} & 88 & 86   & 162,8 & 600,4 & 2727,4 & 16458,2 \\ \hline
			{\bf SAT-4}     & 76,2	& 112,6	& 239,4	& 844 &	4571,8	& 25017 \\ \hline
			{\bf UNSAT-4} & 71,2	& 76,4	& 175,6	& 566,4	& 2687,8 & 16355,6 \\ \hline
			{\bf SAT-8}     & 82,6 & 163,8 & 246,4 & 942,6 & 6199,4 & 21004 \\ \hline
			{\bf UNSAT-8} & 74,4	& 89,8	& 173 & 544	& 2876,8 & 16327,8 \\ \hline
			
		\end{tabular}
	\end{center}
	\caption{Solving time results (time in ms).}\label{tbl:solving-times}
\end{table}

	In general, the validator showed good efficiency, solving 2 annotations in less than 100 ms, while a set with 64 interrelated annotations
	took about 25 seconds. We believe these are acceptable times, which for sets of about 10 to 20 interrelated annotations amounts to
	analysis times of less than or around 1 second. For example, JPA contains 8 annotations for classes and 11 for fields.
	Interestingly detecting unsatisfiability is normally quicker than producing an example. Also, it can be observed that
	the number of annotations has a more significant impact on the solving time than the number of constraints per annotation. Even in some
	cases, adding more constraints resulted in lower solving times (as the search space becomes smaller). 

\subsection{Threats to validity}

		The generalizability of the results presented in the JPA use case should be dealt with care, given that the evaluation of the expressiveness has been done over a small, albeit frequent, set of annotations.
	However, a huge coverage of conceptual constraints behind a set of dependent annotations is shown within this simple example. Note also that the definition of the annotation types in Listing \ref{jpa_cod} is a simplification of the original ones, since neither all the interacting annotations from the JPA framework have been considered, nor all their specification has been translated into Ann. However, we believe we have chosen a subset of the more representative and frequent constraints a JPA developer comes up with, big enough to motivate the usefulness of Ann in a real scenario.

	Regarding the efficiency evaluation, the experiment is synthetic, and therefore may not emulate well the constraints within real-world
	annotations. However, as the bigger impact of solving time is in the number of annotations, our experiments went to a high number of
	annotations (64), showing acceptable times. 

\section{Conclusions and future work}\label{conclusions}
Ann makes possible the effective design of Java annotations by improving their native syntactical support and
allowing the expression of integrity constraints both related to an annotation type and within a set of annotations.
Thanks to the code generator, the approach can be perfectly integrated with existing Java projects. 
Moreover, with the use of annotation processors all the integrity constraints described with the DSL are 
checked at compile time, which improves both usability and efficiency. This is because it is not necessary to
execute the application in order to know whether the annotations are being correctly used, 
hence saving much time and effort for developers. By interacting with a constraint solver,
it provides feedback to the designer of the annotations while they are being constructed, in the event
of inconsistencies.
 
Concerning future work, a large range of possibilities is available given the flexibility
that a DSL provides. 
First, improvements can be done at the tool level, for a smoother integration with the model finder. 
Automated support for finding the reasons for a constraint conflict, or signalling 
redundant constraints is an interesting line of research. We will also consider an empirical
evaluation of Ann with Java programmers.
As seen in Section \ref{related}, the meta-model of Java annotations can be still
improved and expanded to improve its harmony with the rest of Java elements, like, for example, 
its conciliation with object-oriented principles such as composition, inheritance and polymorphism,
which might help to make cleaner the design of a set of annotations.
We also plan to provide support for the new of Java 8 concerning annotations, like
new targets for annotations (any type use). Among other considerations, this may imply using a more
complete Java meta-model for the analysis, like those provided by JaMoPP\footnote{\url{http://www.jamopp.org/index.php/JaMoPP}} 
or Modisco\footnote{\url{https://eclipse.org/MoDisco/}}.

At present two basic types of constraints are considered in Ann (requirements and prohibitions), which
are enough to express common integrity constraints as it has been seen in Section \ref{usecase}. However, 
further experimentation could reveal new constraint types or combinations, which could be added 
to the DSL in the future, given the flexibility that a meta-model provides. Another line of work is the reverse engineering of annotation constraints from the analysis of annotated Java programs.

\section*{\bf Acknowledgements} We would like to thank the reviewers for their detailed comments, which
helped us in improving a previous version of this paper.
This work has been partially supported by the Spanish Ministry of Economy and Competitivity 
with project FLEXOR (TIN2014-52129-R) and the Community of Madrid with project SICOMORO-CM (S2013/ICE-3006).

\bibliographystyle{abbrv}
\bibliography{article}  

\begin{appendices}

\section{Textual concrete syntax of Ann}\label{an:sint}
This appendix includes the complete textual concrete syntax of the Ann DSL in \textit{Extended Backus-Naur Form}.

\subsection{Attributes}
\begin{lstlisting}[language=EBNF, style=EBNF]
Attribute ::=	ClassAtt | StringAtt | ExternalAtt  
			| IntAtt | LongAtt | ShortAtt | FloatAtt | DoubleAtt 
			| ByteAtt | CharAtt | BooleanAtt;

ClassAtt ::= "Class" ("[]")? ID ("=" ClassDefault)?;

StringAtt ::= "String" ("[]")? ID ("=" STRING)?;

ExternalAtt ::= ID ("[]")? ID ("=" (EnumDefault | AnnDefault))?;

IntAtt ::= "int" ("[]")? ID ("=" INT)?;

LongAtt ::= "long" ("[]")? ID ("=" INT)?;

ShortAtt ::= "short" ("[]")? ID ("=" INT)?;

FloatAtt ::= "float" ("[]")? ID ("=" FLOAT)?;

DoubleAtt ::= "double" ("[]")? ID ("=" FLOAT)?;

CharAtt ::= "char" ("[]")? ID ("=" CHAR)?;

BooleanAtt ::= "boolean" ("[]")? ID ("=" BOOLEAN)?;

ByteAtt ::= "byte" ("[]")? ID ("=" BYTE)?;

AnnDefault ::= AnnID
			| AnnID "(" AnnValue ")"
			| AnnID "(" KeyValue ("," KeyValue)* );

ClassDefault ::= ID ".class";

EnumDefault ::= ID "." ID;

AnnID ::= "@" ID;

KeyValue ::= ID "=" AnnValue;

AnnValue ::= AnnArray | AnnBasicValue;

AnnArray ::= "{" "}"
			|  "{" AnnBasicValue ("," AnnBasicValue)* "}";

AnnBasicValue :: EnumDefault | AnnDefault  
			| FLOAT | INT | BOOLEAN | CHAR | BYTE | STRING;
\end{lstlisting}

\subsection{Annotations}

\begin{lstlisting}[language=EBNF, style=EBNF]
Annotation ::= Retention? "annotation" ID
	{
		(Attribute ";")*
		Constraints?
	};

Constraints ::= (Require | Forbid)+;

Retention ::= "runtime" | "class" | "source";
\end{lstlisting}

\subsection{Constraints}

\begin{lstlisting}[language=EBNF, style=EBNF]
Forbid ::= "forbid" Statement ("and" Statement)* ";"; 
			| "at" TargetType : "forbid" Statement ("and" Statement)* ";";

Require ::= "require" Statement ("or" Statement)* ";";
			| "at" TargetType : "require" "all"? Statement ("or" Statement)* ";";

Statement ::= AnnID	| TgtStatement;

TgtStatement ::= AnnID? Modifiers TargetType;

Modifiers ::= VisibMod? & "final"? & "abstract"? & "static"?;
	
VisibMod ::= "public" | "private" | "protected" | "package";

TargetType ::= "interface" | "class" | "annotation" 
			| "method" | "field" | "constructor" | "enum";
\end{lstlisting}

\section{USE model and OCL constraints generated for the JPA annotations}\label{an:ocl}

This appendix includes the generated USE model and OCL constraints for the JPA annotations
in Listing~\ref{jpa_cod}. Please note that, with respect to the meta-model in Figure~\ref{jmm}, some
classes have been prefixed with ``Java'' (e.g., \code{JavaClass}, \code{JavaAnnotation}) to avoid name clashes with
reserved words in USE.

\begin{lstlisting}[language=OCLUSE, style=JavaColors, label=OCLExample]
class Entity < JavaAnnotation
attributes
name : String
	constraints
    inv redefs : self.target.isUndefined()
    
    inv at_class__require_public_constructor_or_protected_constructor: 
    	self.targetEntityClass->notEmpty() implies (
		(self.targetEntityClass.methods->exists(e | 
			e.isConstructor = true and e.visibility = #public) or 
		(self.targetEntityClass.methods->exists(e | 
			e.isConstructor = true and e.visibility = #protected))
    	)
    inv at_class__require_annId_method_or_annId_field_or_annEmbeddedId_method_or_annEmbeddedId_field: 
    	self.targetEntityClass->notEmpty() implies (
			(self.targetEntityClass.methods->exists(e | 
				e.annotationsId->notEmpty())) or 
    		(self.targetEntityClass.fields->exists(e | 
				e.annotationsId->notEmpty())) or 
    		(self.targetEntityClass.methods->exists(e | 
				e.annotationsEmbeddedId->notEmpty())) or 
    		(self.targetEntityClass.fields->exists(e | 
				e.annotationsEmbeddedId->notEmpty()))
    	)
    
    inv forbid_final_class: 
    	not (
    		(self.targetEntityClass.isFinal = true)
    	)
     
    inv at_class__forbid_final_method: 
    	self.targetEntityClass->notEmpty() implies not (
    		(self.targetEntityClass.methods->exists(e | e.isFinal = true))
    	)
     
    inv at_class__forbid_annId_method_and_annEmbeddedId_method: 
    	self.targetEntityClass->notEmpty() implies not (
			(self.targetEntityClass.methods->exists(e | 
				e.annotationsId->notEmpty())) and 
    		(self.targetEntityClass.methods->exists(e | 
				e.annotationsEmbeddedId->notEmpty()))
    	)
     
    inv at_class__forbid_annId_field_and_annEmbeddedId_field: 
    	self.targetEntityClass->notEmpty() implies not (
			(self.targetEntityClass.fields->exists(e | 
				e.annotationsId->notEmpty())) and 
    		(self.targetEntityClass.fields->exists(e | 
				e.annotationsEmbeddedId->notEmpty()))
    	)
     
end
 
class Id < JavaAnnotation
	constraints
    inv redefs : self.target.isUndefined()
    
    inv at_field__require_annEntity_class: 
    	self.targetIdField->notEmpty() implies (
			Entity.allInstances()->exists(e | 
				e.targetEntityClass = self.targetIdField.owner)
    	)
    inv at_method__require_annEntity_class: 
    	self.targetIdMethod->notEmpty() implies (
    			Entity.allInstances()->exists(e | 
				e.targetEntityClass = self.targetIdMethod.owner)    		
    	)
    
end
 
class IdClass < JavaAnnotation
attributes
value : JavaClass
	constraints
    inv redefs : self.target.isUndefined()
    
    inv require_annEntity_class: 
    		(self.targetIdClassClass.annotationsEntity->notEmpty())
    
end
 
class Embeddable < JavaAnnotation
	constraints
    inv redefs : self.target.isUndefined()   
    
    inv at_class__forbid_annId_method: 
    	self.targetEmbeddableClass->notEmpty() implies not (
    		(self.targetEmbeddableClass.methods->exists(e | 
				e.annotationsId->notEmpty()))
    	)
     
    inv at_class__forbid_annId_field: 
    	self.targetEmbeddableClass->notEmpty() implies not (
    		(self.targetEmbeddableClass.fields->exists(e | 
				e.annotationsId->notEmpty()))
    	)
     
end
 
class EmbeddedId < JavaAnnotation
	constraints
    inv redefs : self.target.isUndefined()
    
    inv at_field__require_annEntity_class: 
    	self.targetEmbeddedIdField->notEmpty() implies (
    		Entity.allInstances->exists( e | 
				e.targetEntityClass = self.targetEmbeddedIdField.owner )    		
    	)
    inv at_method__require_annEntity_class: 
    	self.targetEmbeddedIdMethod->notEmpty() implies (
		    Entity.allInstances->exists( e |
			    e.targetEntityClass = self.targetEmbeddedIdMethod.owner )    		
    	)
    
end
 

association Entity_target_class between
	JavaClass [1..1] role targetEntityClass
	Entity [0..*] role annotationsEntity
end
 
association Id_target_method between
	JavaMethod [0..1] role targetIdMethod
	Id [0..*] role annotationsId
end
 
association Id_target_field between
	JavaField [0..1] role targetIdField
	Id [0..*] role annotationsId
end
 
association IdClass_target_annEntity_class between
	JavaClass [1..1] role targetIdClassClass
	IdClass [0..*] role annotationsIdClass
end
 
association Embeddable_target_class between
	JavaClass [1..1] role targetEmbeddableClass
	Embeddable [0..*] role annotationsEmbeddable
end
 
association EmbeddedId_target_method between
	JavaMethod [0..1] role targetEmbeddedIdMethod
	EmbeddedId [0..*] role annotationsEmbeddedId
end
 
association EmbeddedId_target_field between
	JavaField [0..1] role targetEmbeddedIdField
	EmbeddedId [0..*] role annotationsEmbeddedId
end
\end{lstlisting}

\end{appendices}

\end{document}